\documentclass[fleqn,12pt]{wlscirep}
\usepackage[utf8]{inputenc}
\usepackage[T1]{fontenc}
\usepackage{graphicx}
\usepackage{dcolumn}
\usepackage{bm}
\usepackage{graphics}
\usepackage{subfigure}
\usepackage{graphicx}
\usepackage{epsfig}
\usepackage{epstopdf}
\usepackage{xcolor}
\usepackage{cite}

\DeclareUnicodeCharacter{2212}{-}

\bibstyle{naturemag-doi}

\title{Probing embedded topological modes in bulk-like GeTe-Sb$_2$Te$_3$ heterostructures}

\author[1,2,*]{Hisao Nakamura}
\author[3,4,*]{Johannes Hofmann}
\author[5]{Nobuki Inoue}
\author[6]{Sebastian Koelling}
\author[6]{Paul M. Koenraad}
\author[7]{Gregor Mussler}
\author[7]{Detlev Gr{\"u}tzmacher}
\author[8,*]{Vijay Narayan}

\affil[1]{CD-FMat, National Institute of Advanced Industrial Science and Technology (AIST), 1-1-1 Umezono, Tsukuba Central 2, Tsukuba, Japan}
\affil[3]{Department of Applied Mathematics and Theoretical Physics, University of Cambridge, Centre for Mathematical Sciences, Cambridge CB3 0WA, United Kingdom}
\affil[2]{Department of Materials Science and Metallurgy, University of Cambridge, 27 Charles Babbage Road, Cambridge CB3 0FS, United Kingdom}
\affil[4]{TCM Group, Cavendish Laboratory, University of Cambridge, Cambridge CB3 0HE, United Kingdom}
\affil[5]{RIKEN Center for Computational Science, 7-1-26 Minatojima-minami, Cyuo-ku, Kobe, Hyogo 650-0047, Japan}
\affil[6]{Eindhoven University of Technology, 5600 MB Eindhoven, The Netherlands}
\affil[7]{Peter Grünberg Institute (PGI-9), Forschungszentrum Jülich, 52425 Jülich, Germany}
\affil[8]{Department of Physics, University of Cambridge, J. J. Thomson Avenue, Cambridge CB3 0HE}

\affil[*]{hs-nakamura@aist.go.jp, jbh38@cam.ac.uk, vn237@cam.ac.uk}

\begin{abstract}
The interface between topological and normal insulators hosts metallic states that appear due to the change in band topology. While these topological states at a surface, i.e., a topological insulator-air/vacuum interface, have been studied intensely, topological states at a solid-solid interface have been less explored. Here we combine experiment and theory to study such \textit{embedded} topological states (ETSs) in heterostructures of GeTe (normal insulator) and Sb$_2$Te$_3$ (topological insulator). We analyse their dependence on the interface and their confinement characteristics. To characterise the heterostructures, we evaluate the GeTe-Sb$_2$Te$_3$ band offset using X-ray photoemission spectroscopy, and chart the elemental composition using atom probe tomography. We then use first-principles to independently calculate the band offset and also parametrise the band structure within a four-band continuum model. Our analysis reveals, strikingly, that under realistic conditions, the interfacial topological modes are delocalised over many lattice spacings. Interestingly, the first-principles calculations indicate that the ETSs are relatively robust to disorder and this may have practical ramifications. Our study provides insights into how to manipulate topological modes in heterostructures and also provides a basis for recent experimental findings [Nguyen \textit{et al.}, Sci. Rep. \textbf{6}, 27716 (2016)] where ETSs were seen to couple over large distances.

\end{abstract}

\begin{document}

\flushbottom
\maketitle
\thispagestyle{empty}

\section{INTRODUCTION}

Topological surface states have been studied intensively for over a decade, and in particular surface spectroscopic methods such as angle-resolved photoemission spectroscopy (ARPES) have been instrumental in visualising their properties. However, topological modes form whenever there is a interface between and topological and ordinary insulator, and not exclusively when the latter is air/vacuum. Embedded topological states (ETSs), i.e., those that form at the interface between topological and non-topological solid materials, offer interesting possibilities such as the controlled rendering of other topological phases~\cite{burkov11}, whilst also providing a simple means to shield topological states from environmental perturbations (humidity, air pressure, dust etc.). Clearly, ETSs cannot be probed using ARPES, and in order to understand and subsequently manipulate them, one needs to devise alternate methods. 

The GeTe-Sb$_2$Te$_3$ system is a TI-NI system that offers clear advantages towards studying ETSs. Here, GeTe is a normal insulator (NI)~\cite{tsu68} that becomes superconducting below $\approx~1$~K~\cite{hein64,narayan16,narayan19}, and Sb$_2$Te$_3$ is a topological insulator (TI)~\cite{hiseh09,zhang09}. From a materials perspective, the GeTe-Sb$_2$Te$_3$ system is well-known due to its phase-change properties~\cite{raoux14,kato06,yamada87,simpson11,chen86,inoue19}, ferroelectric characteristics~\cite{tominaga14,tominaga15}, and potential thermoelectric properties~\cite{ibarra18}. On the other hand, superlattices of alternating Sb$_2$Te$_3$ and GeTe layers are known to have a non-trivial band topology that is governed by (i) the coupling of topological modes~\cite{burkov11, halasz12} that appear at each Sb$_2$Te$_3$-GeTe interface in the superlattice, and (ii) the physical intermixing of adjacent layers which is significant owing to the strong chemical affinity between the materials. However, to date the majority of investigations, theoretical and experimental, focuses on monolayer/few monolayer superlattice units~\cite{tominaga14,tominaga15,sa12,nakamura17,bang16,qian16} for which intermixing easily destroys the layered heterostructure~\cite{Momand2015}. In contrast, we will be interested in relatively less explored structures where the individual GeTe and Sb$_2$Te$_3$ layers are more bulk-like, and consequently, in which individual GeTe-rich and Sb$_2$Te$_3$-rich regions are well-defined. This approach allows us to study individual ETSs which are hard to delineate in superlattices. 

%The superlattice system consisting of $L$ GST units, labeled as [(Sb$_2$Te$_3$)$_M$(GeTe)$_N$]$_L$, is  overall a topological or non-topological insulator depending on the relative layer thickness denoted by the indices $M$ and $N$~\cite{tominaga14, halasz12}.

%Here, Sb$_2$Te$_3$ is a topological insulator (TI)~\cite{hiseh09,zhang09} and GeTe is a normal insulator (NI)~\cite{tsu68}, which becomes superconducting below $1 K$~\cite{hein64,narayan16,narayan19}. 

%There are two mechanisms that contribute to the character of the band structure: (i) the physical intermixing of adjacent layers, and (ii) the coupling of topological modes~\cite{burkov11, halasz12} that appear at the interface of Sb$_2$Te$_3$ and GeTe (SG interface) in the superlattice. Importantly, it is known that the strong chemical affinity between GeTe and Sb$_2$Te$_3$ induces significant intermixing at the interface where the topological state is expected.
%These two-dimensional modes are localised on the edge of (Sb$_2$Te$_3$)$_M$ layer and have a linear Dirac dispersion with a helical spin structure.  More strictly, the overall phase is governed by the competing hybridization of topological modes of the top and bottom edge of the (Sb$_2$Te$_3$)$_M$ layer (intralayer coupling) and those of the neighboring layers through (GeTe)$_N$ layer (interlayer coupling). 

In this manuscript we present a comprehensive analysis of the band structure of bulk-like GeTe-Sb$_2$Te$_3$ heterostructures and the ETSs that form at the heterointerface. This treatment derives from detailed materials characterisation of the heterostructures including X-ray photoemission spectroscopy (XPS) to evaluate the band offset, and Atom Probe Tomography (APT) to visualise the spatial atomic distribution. We first calculate the band-offset between GeTe and Sb$_2$Te$_3$ from first principles and verify our result against the XPS measurements. We then develop a continuum model of GeTe-Sb$_2$Te$_3$ heterostructures based on parameters extracted from the first-principles calculation. The continuum model shows that the physical structure at the Sb$_2$Te$_3$-GeTe interface strongly influences the extent over which ETS are localised and, therefore, in mediating inter-ETS coupling in multi-layer heterostructures. More specifically, we find that under conditions of a sharp, well-defined interface, interactions between topological modes separated by more than a few nm is relatively weak, but under more realistic conditions where the interface has some degree of intermixing, topological modes separated by as much as $10$~nm couple and develop a gap of several meV. Finally, we use first-principles calculations to understand the impact of chemical disorder on the ETSs and establish that these show a striking degree of resilience to structural disorder arising from the intermixing.

%This strongly supports the findings of Nguyen~\textit{et al.}~\cite{nguyen16}
%In this context, a recent finding by some of the authors showed, unexpectedly, that in heterostructures of bulk-like Sb$_2$Te$_3$ and GeTe, the net number of topological modes is controlled by the GeTe layer even in a range where $M$, $N \gg 10$~\cite{nguyen16}. In Ref.~\cite{nguyen16}, the low-temperature magneto-transport of a Sb$_2$Te$_3$-GeTe-Sb$_2$Te$_3$ tri-layer structure was measured and the number of two-dimensional (2D) modes was inferred from the weak anti-localization characteristic~\cite{hikami80,backes17,backes19}. Whereas one might expect that each TI/NI interface should host a topological mode, it was observed that for samples with a $15$~nm-thick GeTe layer, the number of 2D modes was two fewer than expected. It was suggested that in a Sb$_2$Te$_3$-GeTe-Sb$_2$Te$_3$ (SGS) tri-layer, the topological modes flanking the GeTe layer hybridised even when separated by up to 15~nm and developed a gap of approximately $30$~K. In order to understand this SGS tri-layer, i.e., TI/NI/TI tri-layer system, the role of band offsets at the SG interface, atomistic structure, and electronic structure of the junction have to be clarified.

This paper is structured as follows: In Sec.~\ref{sec:2}, we present experimental measurements of the band offset between Sb$_2$Te$_3$ and GeTe on molecular-beam-epitaxy (MBE)-grown samples using X-ray photoemission spectroscopy (XPS). We also present atom probe tomography (APT) of a Sb$_2$Te$_3$-GeTe-Sb$_2$Te$_3$ (SGS) heterostructure which shows well-defined Sb$_2$Te$_3$ and GeTe regions separated by narrow but finite intermixed regions. In Sec.~\ref{sec:2}.2 we use first-principles calculations to obtain an independent estimate of the band offset at the SG interface for a range of experimentally relevant microstructures. Our estimates compare favorably with the XPS data, thereby validating the first-principles result. Next, in Sec.~\ref{sec:3}, we develop a continuum model of an SGS tri-layer within a four-band framework, the parameters of which are obtained from our first-principles calculations. The continuum model can be used to study large systems that would be computationally too expensive to study directly using first-principles methods. Finally, in Sec.~\ref{sec:IV}, we analyze the robustness of topological modes to chemical interactions and examine how the microscopic structure of the interfacial region may impact the interlayer coupling of topological modes.

\section{BAND OFFSET}\label{sec:2}

In this section, we discuss the band offset between the GeTe and Sb$_2$Te$_3$ layers in the GST heterostructure. First, in Sec.~\ref{sec:experiment}, we present experimental measurements of the band offset at the interface of GeTe and Sb$_2$Te$_3$ using XPS depth profiles. In Sec.~\ref{sec:II2}, we then use first-principles calculations to evaluate the bulk crystal structure of GeTe and Sb$_2$Te$_3$, from which we obtain an independent evaluation of the band offset, which agrees with the experimental results. The agreement between experiment and theory validates the theoretical calculation, and provides a basis for the continuum model developed in Section~\ref{continuum} to describe larger systems.

\subsection{EXPERIMENTAL EVALUATION}\label{sec:experiment}

We use XPS measurements to determine the band offset between bulk GeTe and bulk Sb$_2$Te$_3$. Following Refs.~\cite{kim07,kraut80,fang11}, this is done by evaluating the difference of the core electron energy levels in bulk samples, and comparing this to the difference in the core energy levels in a heterostructure. Conventionally, this would require XPS spectra of three separate samples: a bulk GeTe film, a bulk Sb$_2$Te$_3$ film, and a heterostructure of the two in which the top layer is sufficiently thin ($\sim 5$~nm) that the X-rays can penetrate it fully and sample both materials. However, such a procedure will have unknown systematic errors when considering Sb$_2$Te$_3$-GeTe heterostructures as the two compounds have a strong chemical affinity and will undergo significant intermixing, especially in the vicinity of the interface. Here, we obtain instead a ``depth profile'' of a single GeTe-Sb$_2$Te$_3$ heterostructure where XPS spectra are taken between successive Ar-ion etches of the sample, which successively remove the top layers of the heterostructure. The measurements are continued for the entire depth of the sample, i.e., until the Ar-ion etch fully depletes the material. This approach eliminates variations due to the different growth conditions for separate samples. The samples considered here are MBE-grown heterostructures in which the GeTe (top) layer has a thickness of 11~nm and the Sb$_2$Te$_3$ layer is 25~nm thick. The samples are grown on a Si(111) substrate as described in Ref.~\cite{nguyen16}.

%++++++++++++++++++++++++++++++++++++++++
\begin{figure}[t]
\begin{center}
\subfigure[\qquad]{\scalebox{0.23}{\includegraphics{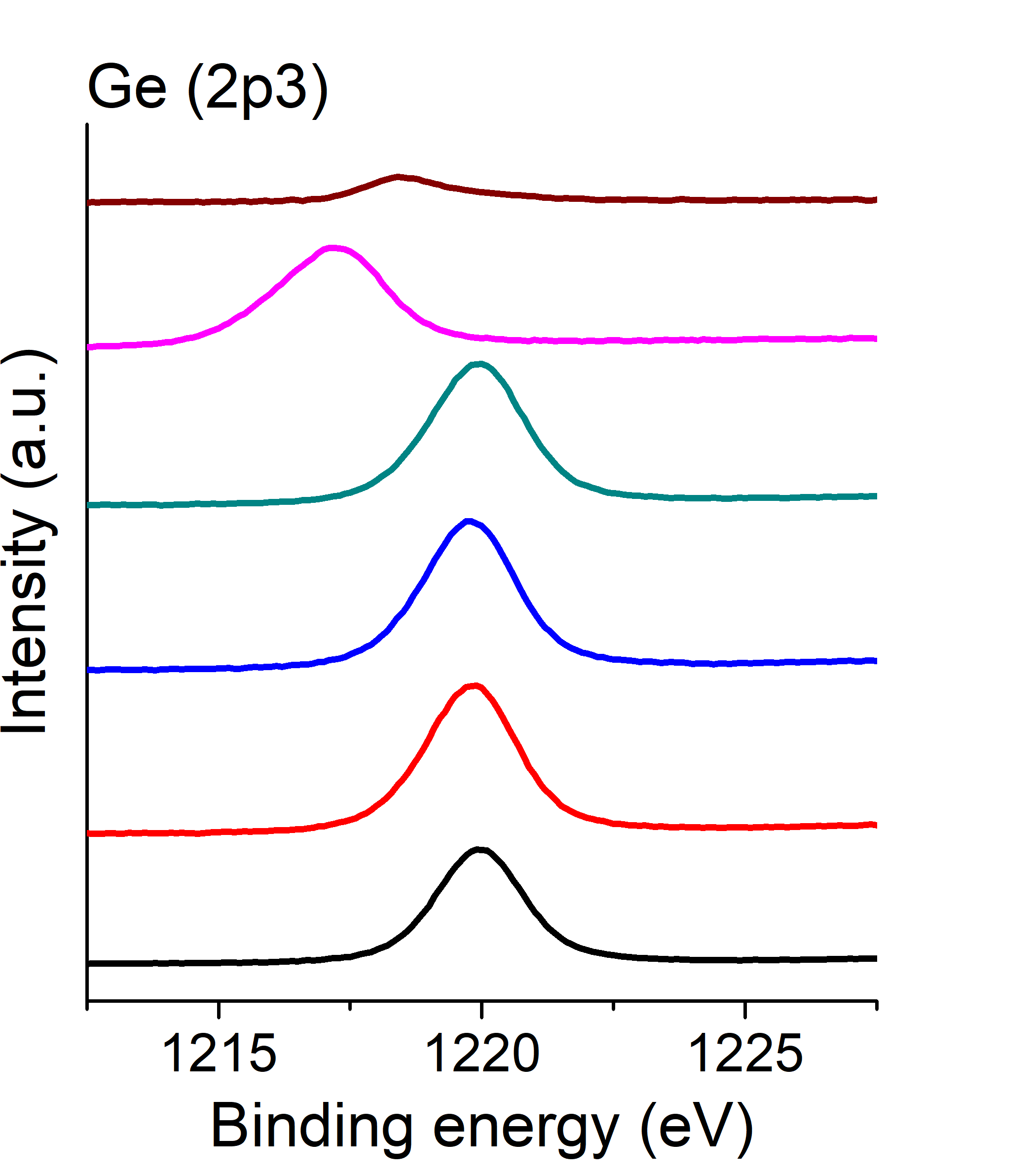}\label{fig:1a}}}\quad
\subfigure[\qquad]{\scalebox{0.23}{\includegraphics{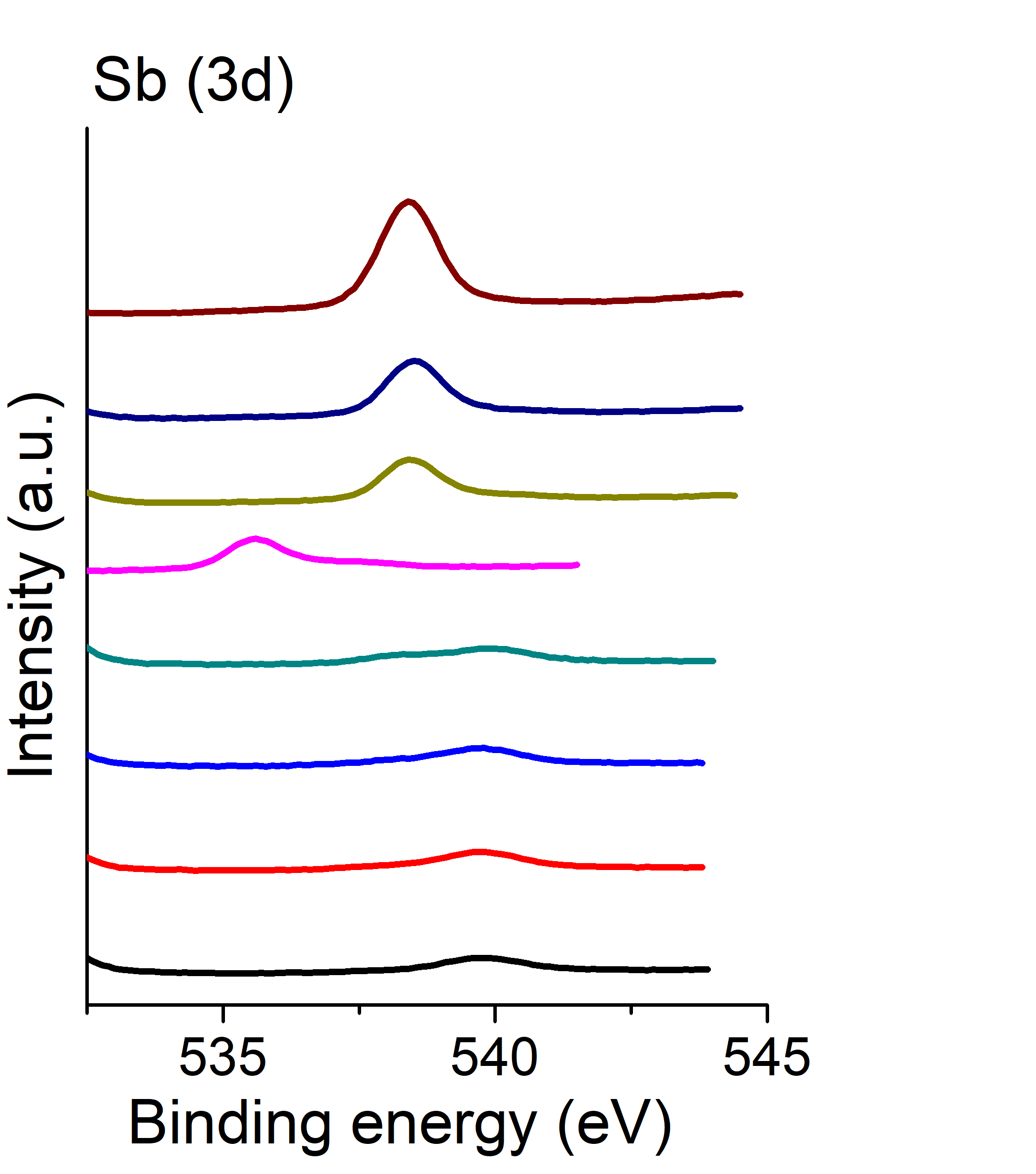}\label{fig:1b}}}\quad
\subfigure[\qquad]{\scalebox{0.23}{\includegraphics{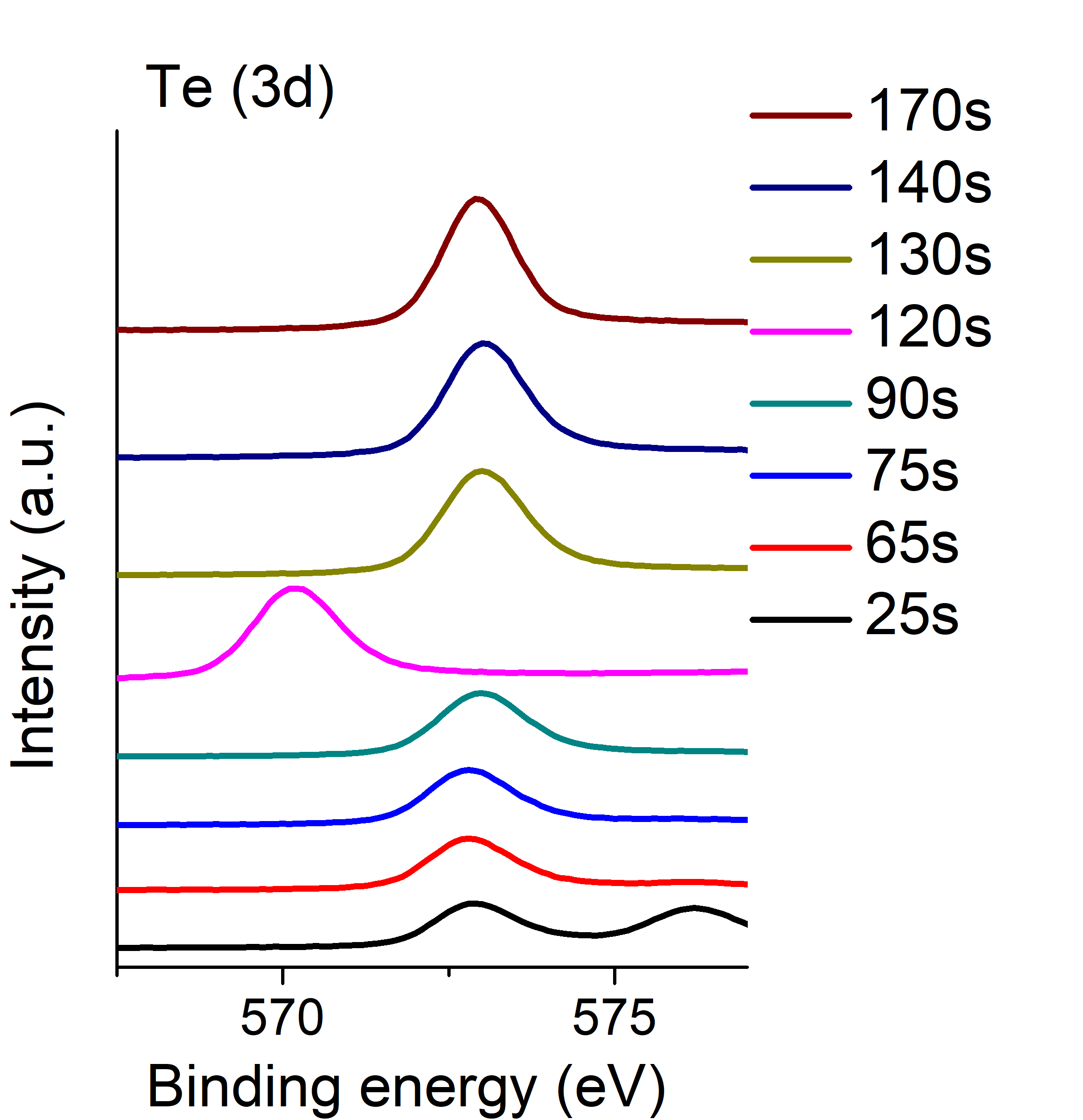}\label{fig:1c}}}
\caption{XPS spectra of the GeTe-Sb$_2$Te$_3$ heterostructure after consecutive ion milling steps for three different energy ranges near the (a) Ge(2p3), (b) Sb(3d), and (c) Te(3d) peaks. Spectra are obtained for eight different etch times of (bottom to top) $t=25, 65, 75, 90, 120, 130, 140$, and $170$s, and we include an arbitrary offset between spectra to guide the eye. Panel (a) shows an initial pronounced Ge(2p3) peak at $1220$eV that nearly vanishes after $170$s of milling, indicating that the GeTe is fully eroded at that time. Consistent with this, panel (b) shows a clear peak at $540$eV, corresponding to the Sb (3d) transition, that emerges after $120$s of etching, indicating the absence of Sb in the top layers of the original unetched sample. Panel (c) shows a consistent peak at $573$eV corresponding to the Te(3d) transition, indicating that the Te is present throughout the heterostructure. Panels (a) and (b) also show residual Sb(Ge) peaks in the Ge(Sb)-rich regions, which suggests an intermixing between the GeTe and Sb$_2$Te$_3$ layers.}
\label{fig:1}
\end{center}
\end{figure}
%++++++++++++++++++++++++++++++++++++++++

%++++++++++++++++++++++++++++++++++++++++
\begin{figure}[t]
\begin{center}
\subfigure[]{\scalebox{0.17}{\includegraphics{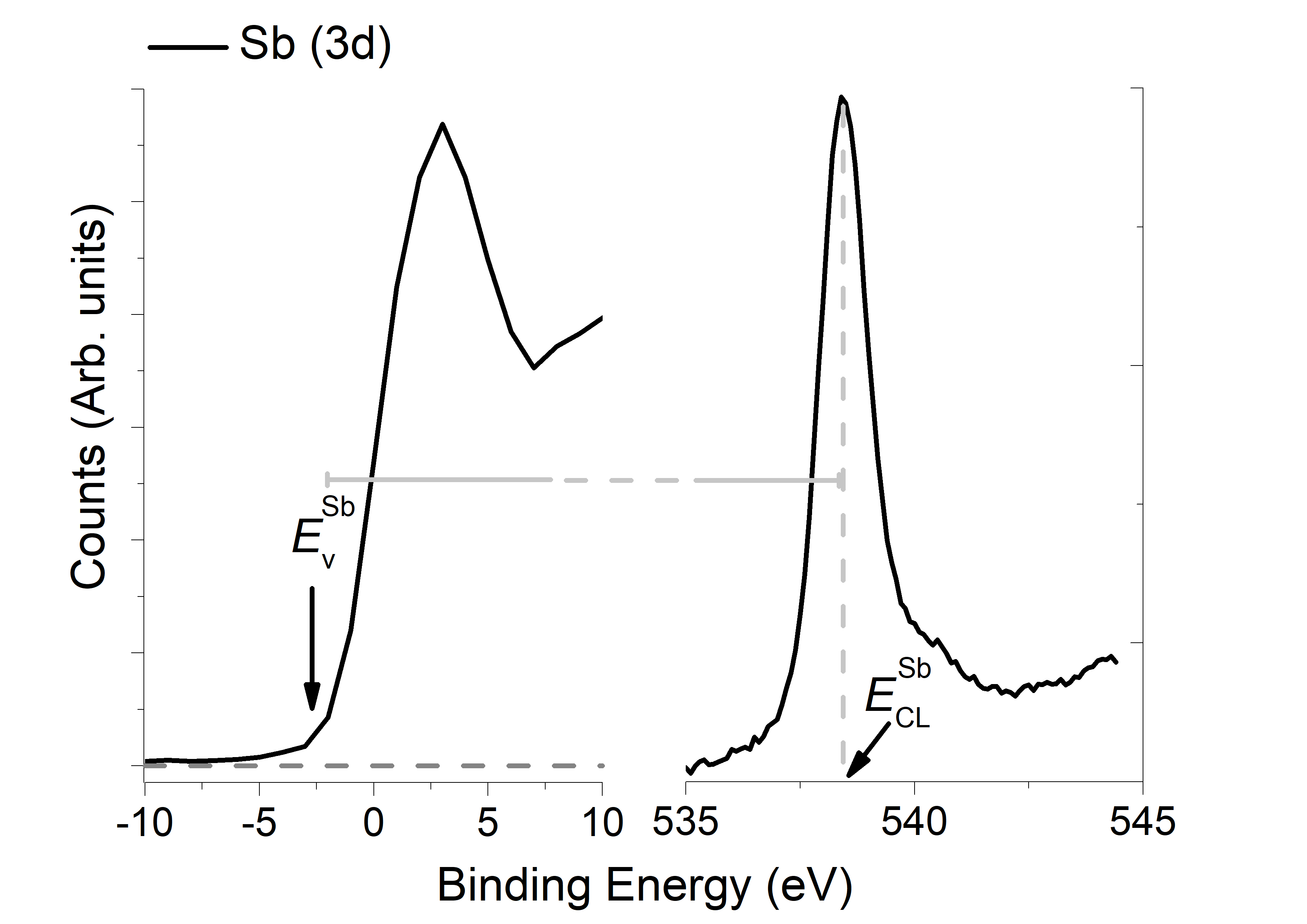}\label{fig:2a}}}
\subfigure[]{\scalebox{0.17}{\includegraphics{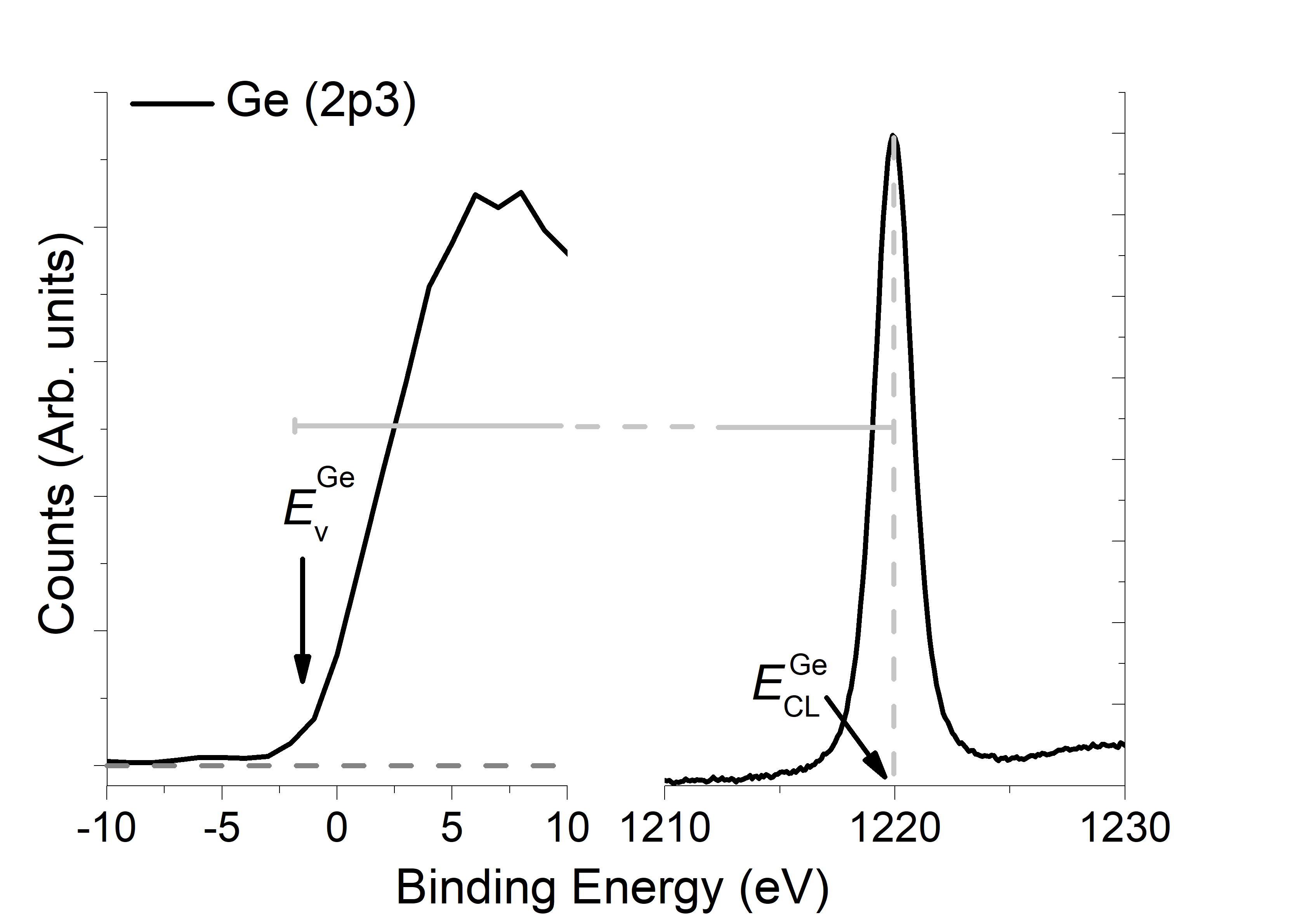}\label{fig:2b}}}
\subfigure[]{\scalebox{0.17}{\includegraphics{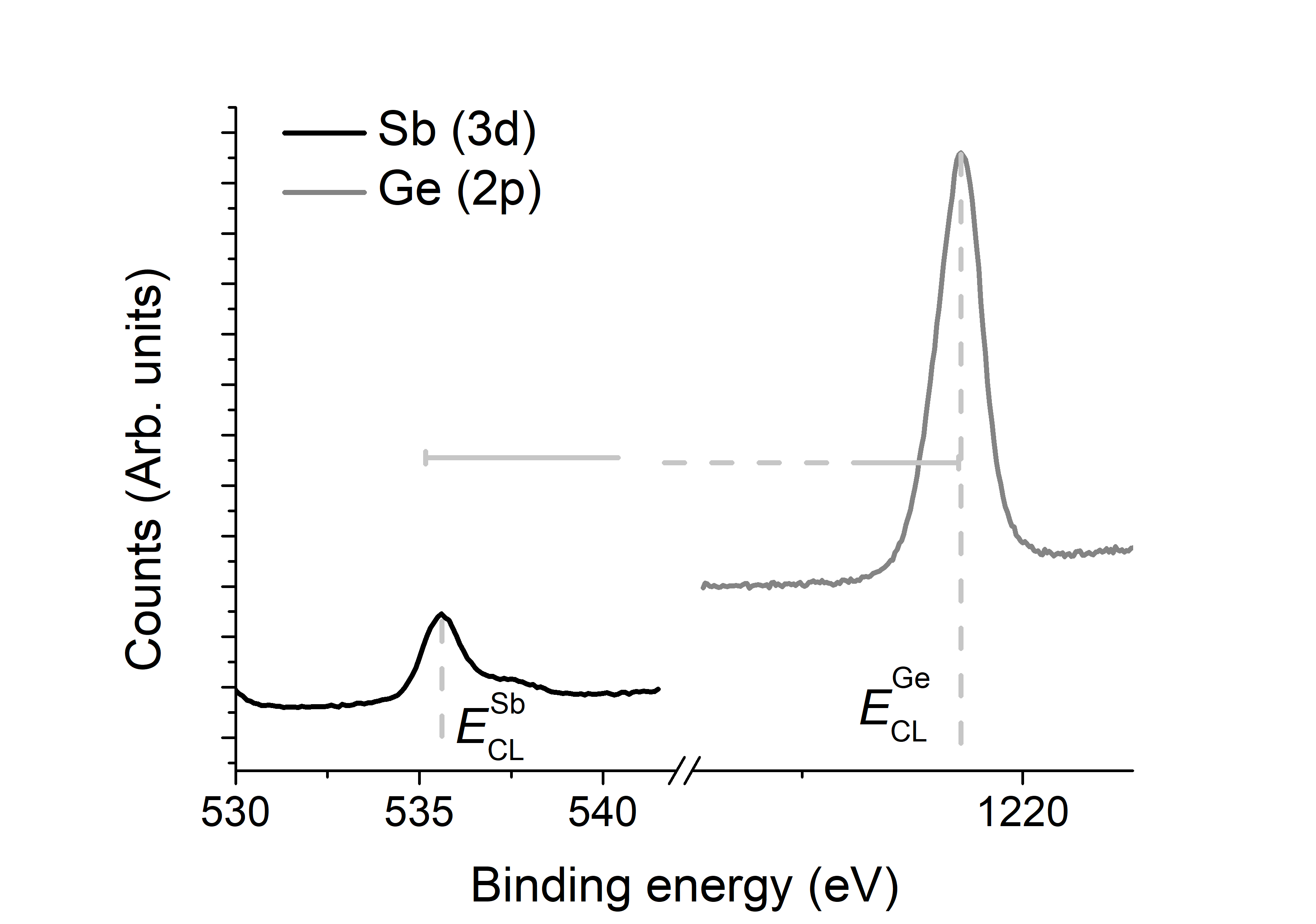}\label{fig:2c}}}
\caption{Absolute position of the (a) Sb(3d) and (b) Ge(2p3) core levels with respect to the valence band energy $E_v$. (a) shows an XPS spectrum measured in the Ge-rich phase and (b) shows an XPS spectrum in the Sb-rich phase. The valence band energy $E_v$ is defined as the minimum energy at which emissions are observed, i.e., at which the XPS spectrum just becomes non-zero (indicated by arrows in (a) and (b)). This is obtained as the intersection of the binding energy curve with the background level in the respective left panels of Fig.~2(a) and 2(b). (c) XPS spectrum after $120$~s of etching which contains signals from both Ge and Sb core levels, from which the difference in core energy levels is evaluated as shown.}
\label{fig:2}
\end{center}
\end{figure}
%++++++++++++++++++++++++++++++++++++++++

Figure~\ref{fig:1} shows XPS depth profiles for different etching times $t=25, 65, 75, 90, 120, 130, 140$, and $170$s in which the Ge(2p3) transition~[Fig.~\ref{fig:1a}], the Sb(3d) transition~[Fig.~\ref{fig:1b}], and the Te(3d) transition~[Fig.~\ref{fig:1c}] is monitored. As expected, initially there is a pronounced Ge peak which begins to diminish at the same time the Sb peak appears, indicating that the top GeTe layer is completely eroded after $130$s of etching. The Te(3d) transition [Fig.~\ref{fig:1c}] shows little depth dependence, which is expected. The traces taken between $120$s and $140$s show features corresponding to both Ge and Sb, suggesting that the X-rays probe both layers in this range, which points to an intermixing between the layers. These traces also reflect the diffuse nature of interface between the two materials. The band offset $\Delta E_v$ is obtained as described in Refs.~\cite{kim07,kraut80,fang11}:
\begin{equation}
\label{XPS_Eq}
\Delta E_v = (E_{\rm CL}^{Ge} - E_v^{Ge}) - (E_{\rm CL}^{Sb} - E_v^{Sb}) + \Delta E_{\rm CL} .
\end{equation}
Here, the first two terms on the right-hand side represent the difference in energy between the core level (CL) and valence band edge ($E_v$) of Ge and Sb, respectively. These are obtained from the bulk spectra as shown in Figs.~\ref{fig:2a} and~\ref{fig:2b}. The third term is the difference in energy between the core levels of Ge and Sb obtained from the combined spectrum shown in Fig.~\ref{fig:2c}. The result for the band offset is $\Delta E_v = 0.4 \pm 0.2$~eV.

\begin{figure}[t]
\begin{center}
\scalebox{0.35}{\includegraphics{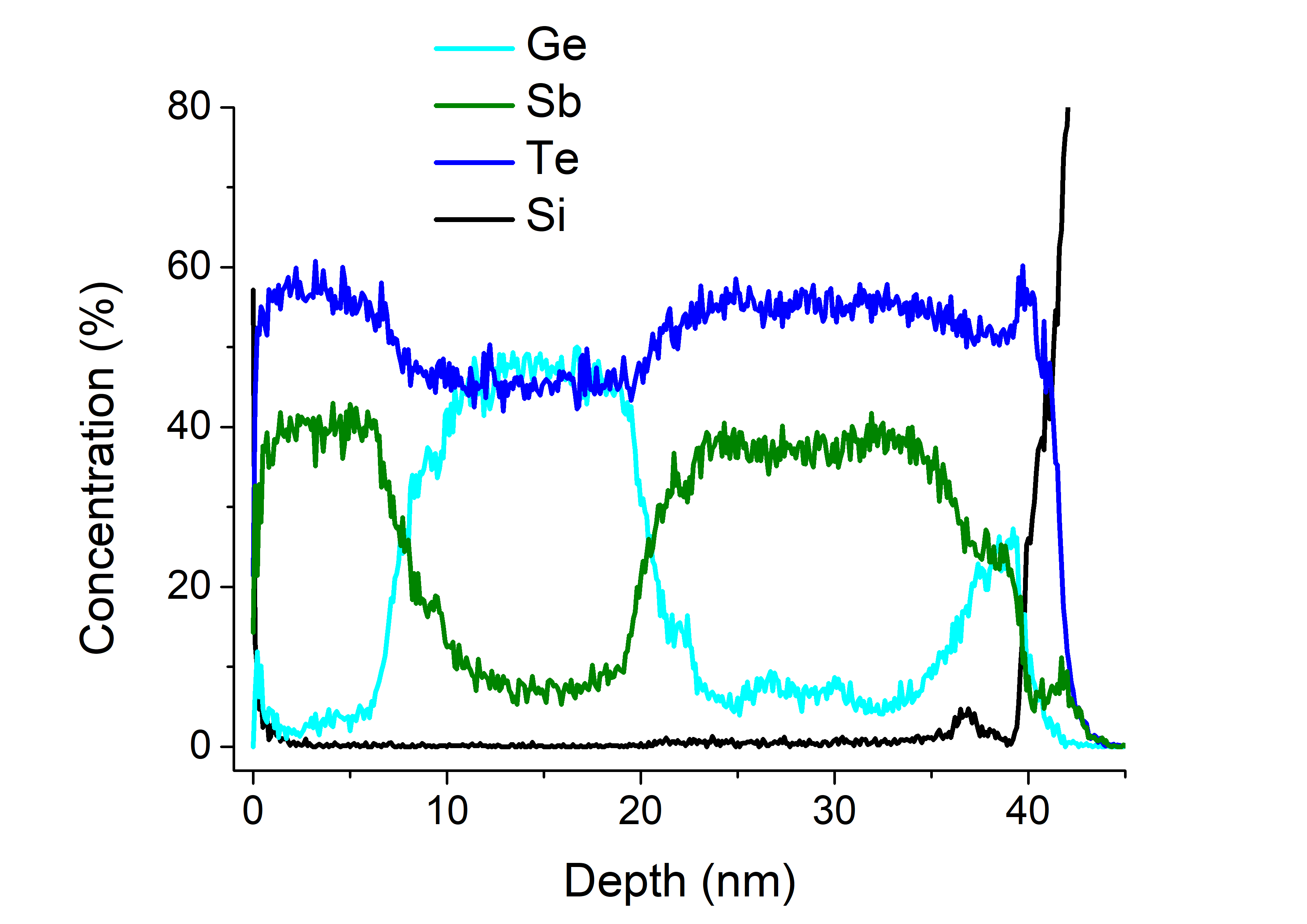}}
\caption{Atom probe tomography (APT) of Sb$_2$Te$_3$/GeTe/Sb$_2$Te$_3$ samples show a depth profile of the concentrations of Ge, Sb and Te through the sample thickness (see Supplementary Material for APT methods~\cite{SOM}). One clearly observes distinct Sb-rich, Ge-rich and Sb-rich regions separated by intermixed regions located at $8$~nm and $22$~nm. Within the Sb (Ge)-rich region there is a {\em relative concentration} of $<~20$\% Ge (Sb) whereas in the interfacial regions (centred at $\approx~8$~nm and $\approx~21$~nm) the relative concentrations of Ge and Sb are equal.}
\label{APT}
\end{center}
\end{figure}

Further evidence for intermixing region is obtained in Fig.~\ref{APT} where we show atom probe tomography (APT) of a Sb$_2$Te$_3$/GeTe/Sb$_2$Te$_3$ sample. APT is based on the evaporation of atoms in the form of ions from a single tip-shaped sample by means of an electric field. During the analyses, ions are projected from the apex of the tip onto a position-sensitive single ion detector~\cite{blavett93} by the electric field. On the basis of the measured positions and the time-of-flight between the tip apex and the detector surface a 3D reconstruction of the analyzed volume is created~\cite{bas95}. Further details on the APT can be found in the Supplementary Material~\cite{SOM}.

%[Blavette, D.; \textit{et al.}, Nature 1993, \textbf{363} 432?435]
%[Bas, P.; \textit{et al.} Appl.Surf. Sci. 1995, \textbf{87?88}, 298?304].

\subsection{FIRST-PRINCIPLES CALCULATIONS}\label{sec:II2}

\begin{table}
\caption{\label{Energy_levels} Fermi energy, valence band maximum (VBM), and conduction band minimum (CBM). All quantities are given in eV. The last row denotes the indirect band gap.
}
\begin{center}
\item[]\begin{tabular}{lllll}
\hline
 & Sb$_2$Te$_3$ (Rh) & Sb$_2$Te$_3$ (Rk) & GeTe (dRk) & GeTe (Rk) \\
\hline
Fermi level & 0.00 & -0.08 & -0.13 & 0.22 \\
VBM & -0.02 & -0.11 & -0.36 & 0.13 \\
CBM & 0.08 & -0.06 & 0.11 & 0.22 \\
Band Gap & 0.10 & 0.05 & 0.47 & 0.09 \\
\hline
\end{tabular}
\end{center}
\end{table}

In this section, we present results for the band offset obtained from electronic structure calculations of bulk Sb$_2$Te$_3$ and GeTe via density functional theory (DFT) and non-equilibrium Green function (NEGF) theory. In our calculations, we use the SIESTA~\cite{soler02} and Smeagol~\cite{rocha06} program packages, details of which can be found in the Supplementary Material~\cite{SOM}. %and adopt the double zeta plus polarization function (DZP) level basis set. Our results are consistent with the experimental findings presented in the previous section. To evaluate the band offset and identify the SG interfacial electronic states, we make use of the non-equilibrium Green’s function technique~\cite{datta97} combined with DFT (NEGF-DFT). Since NEGF-DFT satisfies semi-infinite boundary condition, our results are free from artificial size effects, which are often problematic in slab model calculations. NEGF-DFT calculations were carried out using the Smeagol program package~\cite{rocha06,rungger08}. We adopt an exchange correlation (XC) functional of the van der Waals (vW) correc§tion, DF2~\cite{klime09}, for total energy calculations and use the local density approximation (LDA) to determine the band structure in NEGF-DFT calculations, where spin-orbit (SO) interaction are included.

%At room temperature, bulk GeTe crystallises in a disordered rock salt (dRk) structure and Sb$_2$Te$_3$ crystallises in a rhombohedral layered (Rh) structure~\cite{dasilva08}. Hence, the conventional hexagonal cell can be taken as the unit cell for both, where the (111) direction of the rock salt corresponds to the (0001) in the conventional hexagonal cell. We set the c-axis to the (0001) direction and define the $z$-coordinate along the c-axis. The SG interface plane is then perpendicular to the (0001) direction. In the conventional hexagonal cell, both GeTe~(dRk) and Sb$_2$Te$_3$~(Rh) have a CBA/CBA/CBA/... stacking, where A, B, and C represent monolayers of Sb$_2$Te$_3$ or GeTe. In other words, the bulk unit cell of Sb$_2$Te$_3$ and GeTe consists of three monolayers, which may be denoted as (Sb$_2$Te$_3$)$_3$ and (GeTe)$_3$, respectively~\cite{dasilva08,nonaka00}. The experimental values of the lattice constants are $a_0=b_0=4.26~\AA$ and $c_0=30.75~\AA$ for Sb$_2$Te$_3$~(Rh) and $a_0=b_0=4.17~\AA$ and $c_0=10.90~\AA$ for GeTe~(dRk)~\cite{nonaka00}. Thus, there is a very small lattice mismatch at the SG interface. For simplicity, we fix the lattice constant as $a_0=b_0=4.25~\AA$ for both of Sb$_2$Te$_3$ and GeTe (as well as the SGS tri-layer) in our computational models and then allow all atoms in the cell to relax to their atomic positions. The atomic structures are shown in Figs.~\ref{fig:3a} and~\ref{fig:3b}, respectively.

The band offset is calculated as follows: first, the Fermi level $E_F$ is obtained from DFT calculations of bulk systems that include spin-orbit interactions. Then, we define an extended cell C by taking $1 \times 1 \times 3$ unit cells and apply the self-consistent NEGF-DFT. The left and right sides of C are connected to the bulk semi-infinitely by the self-energy terms, such that our calculations give the Green’s function projected on C. Using the resulting Green’s functions, we analyze the spectral density and evaluate the conduction band minima (CBM) and valence band maxima (VBM). Next, we carried out NEGF-DFT for the same C while the right side of the cell C is now terminated by vacuum. Practically, we took a vaccuum region of $z_{\rm vac} = 15.0~\AA$ in the $z$-direction. Now, we can introduce the unique definition of the Fermi level $E_F^0$ using the vacuum level, i.e., 
\begin{equation}
E_F^0 = E_F - V_H(z=z_{\rm vac}) ,
\end{equation}
where $V_H$ is the Hartree potential averaged over the $xy$ plane. As the last step, we corrected the values of VBM and CBM by Eq. (2), which are denoted as  $E_v^0$ and $E_c^0$, respectively. We applied the above procedures to Sb$_2$Te$_3$~(Rh) and GeTe~(dRk), and as a reference, also to the rock salt (Rk) structures of Sb$_2$Te$_3$ and GeTe which are possible crystal phases representing vacancy states or at high temperature~\cite{sun06,ohyanagi14,disante13}. The value of $E_F^0$ of Sb$_2$Te$_3$~(Rh) is $-4.61$~eV. 

In Table 1, we set $E_F^0$ of Sb$_2$Te$_3$~(Rh) to zero and list the values of the VBM, the CMB and the Fermi level of different structures relative to this. The band offset between Sb$_2$Te$_3$~(Rh) and GeTe~(dRk), given by the difference between the respective VBM is $\Delta E_v \approx 0.36$eV, which agrees well with our experimental value reported in Sec.~2. The validity of the calculations is also confirmed by noting that the calculated band gap of Sb$_2$Te$_3$ is $0.1$eV, which is consistent with its narrow gap, $p$-type semiconductor character. Likewise, the calculated band gap of GeTe~(dRk)$=0.47$eV is close to the experimental value of $0.6$eV~\cite{tsu68}. We note here that our calculated band gap of GeTe (RK/dRK) is slightly lower compared to the experimental results. We attribute this to the disorder considered in the GeTe models which is known to underestimate the bandgap~\cite{dasilva08, chang1966, Esaki1966, Palaz2017}. 
%We note here that the band gap of GeTe~(Rk/dRK) is underestimated by our theoretical calculation, but disorder is known to increase the experimental band gap, which is consistent with our results~\cite{dasilva08}.

The present results suggest the validity of our computational model in order to analyze the topological modes of an SGS tri-layer quantitatively. In the next section, in order to treat large heterostructures beyond the range of numerical DFT simulations, we construct an effective four-band model with parameters derived from our first-principles calculations.

\section{FOUR-BAND CONTINUUM MODEL}\label{sec:3}

\label{continuum}

In the previous section, we have both experimentally and theoretically elucidated the structure of the SG interface. Experimental measurements of the band gap obtained using XPS were shown to be consistent with theoretical results from {\em ab-initio} DFT calculations of a semi-infinite slab structure, which indicates that our computational DFT model is predictive for these systems. The aim of this section is to extend the theoretical model to tri-layer structures and to address the recent experiments by Nguyen~\textit{et al.}~\cite{nguyen16} by considering the qualitative effect of a thick, bulk-like GeTe intermediate layer on the embedded interface states.

While the numerical DFT method is in principle exact, i.e., it will accurately describe the inter- and intralayer coupling as well as the chemical intermixing at the interfaces, modeling very thick bulk-like heterostructures comes with a prohibitive numerical cost. In practice, we are restricted to very thin structures of typically less than ten layers. In order to make contact with the experiments on bulk-like structures of Ref.~\cite{nguyen16}, in this section, we introduce an effective four-band model using parameter values derived from the bulk calculations presented in the previous section. This model allows us to describe tri-layer structures of arbitrary thickness. Indeed, as a main result of this paper, our findings indicate a significant interlayer-coupling of surface states across the GeTe layer, which is consistent with the experiment~\cite{nguyen16}.

The effective four-band model is predictive for inter- and intralayer coupling effects, but it neglects the physical intermixing of the GeTe and Sb$_2$Te$_3$ phases at the interface, i.e., a reconfiguration of atomic positions. In order to take this into account, we consider additionally a model in which the GeTe film is replaced by a Ge$_2$Sb$_2$Te$_5$ (GST225) crystal phase, which is one of the most standard compositions of the GST alloy. We adopt the Kooi structure of GST225~\cite{kooi02}, the crystal structure of which is shown in Fig.~3 of the supplemental material. To further support our effective model, in Sec.~\ref{sec:IV}, we present ab-initio results for thin heterostructures that are consistent with the results obtained by the four-band model.

\begin{table}
\caption{\label{Band_parameters} Band parameters of the four-band continuum model. The parameter definitions are given in Eq.~(\ref{eq:hamiltonian}). Values for Sb$_2$Te$_3$ are taken from Ref.~\cite{zhang09}, parameters for GeTe and GST225 are extracted from a fit to our first-principles results. The Fermi level of Sb$_2$Te$_3$ is set to zero. }
% \begin{indented}
% \item[]
\begin{center}
\begin{tabular}{lllllllllll}
\hline
 & $A_0$ & 	$A_2$	 & $B_0$ & 	$B_2$	 & $C_0$	 & $C_1$ & 	$C_2$	 & $M_0$ & 	$M_1$	 & $M_2$ \\
\hline
Sb$_2$Te$_3$ 	 & 3.40	 & 0.00	 & 0.84	 & 0.00	 & 0.01	 & -12.39	 & -10.78	 & -0.22 & 	19.64	 & 48.51 \\
GeTe 	 & 2.92	 & 0.00	 & 1.46	 & 0.00	 & -0.11	 & -1.00	 & 3.48	 & 0.79	 & -7.05	 & -33.72 \\
GST225 	 & 0.01	 & 0.00	 & 0.00	 & 0.00	 & 0.00	 & -1.96	 & 0.62	 & 0.14 & 	4.69	 & 4.15 \\
\hline
\end{tabular}
\end{center}
% \end{indented}
\label{tab:2}
\end{table}

% \subsection{CONTINUUM MODEL OF HETEROSTRUCTURE}

A simple description of the electronic spectrum in semiconductor heterostructures is obtained using the envelope function formalism~\cite{bastard86}, and details of this calculation are presented in the supplemental material~\cite{SOM}. The formalism describes separate layers in terms of effective bulk band models, which for topological insulators in the Bi$_2$Se$_3$ and Sb$_2$Te$_3$ family capture the band structure near the $\Gamma$ point~\cite{zhang09,liu10}:
\begin{equation}
H = \begin{pmatrix}
\varepsilon({\bf k}) + M({\bf k}) & B(k_z) k_z & 0 & A(k_\parallel) k_- \cr
B(k_z) k_z & \varepsilon({\bf k}) - M({\bf k})& A(k_\parallel) k_- & 0\cr
0 & A(k_\parallel) k_+ & \varepsilon({\bf k}) + M({\bf k})& - B(k_z) k_z \cr
A(k_\parallel) k_+ & 0 & - B(k_z) k_z & \varepsilon({\bf k}) - M({\bf k})
\end{pmatrix} . \label{eq:hamiltonian}
\end{equation}
Here, adopting the notation of Ref.~\cite{liu10}, $\varepsilon({\bf k}) = C_0 + C_1 k_z^2 + C_2 k_\parallel^2$, $M({\bf k}) = M_0 + M_1 k_z^2 + M_2 k_\parallel^2$, $A(k_\parallel) = A_0$, $B(k_z) = B_0$, $k_\parallel^2 = k_x^2 + k_y^2$, and $k_\pm = k_x \pm i k_y$. The interface is aligned in the $xy$-plane, and the $z$-direction is perpendicular to the interface. 

The parameters of Sb$_2$Te$_3$ were derived in Ref.~\cite{zhang09} and are summarised in Tab.~\ref{tab:2}, where we define the zero of the energy scale at the Fermi level of Sb$_2$Te$_3$. For calculations of multilayer structures, we also require parameter values for an effective model of the GeTe phase. Due to band inversion, the band gap of GeTe~(dRk) increases at the $L$ point in the rock salt structure~\cite{sa12,liebmann16}. The four-band model~(\ref{eq:hamiltonian}) is still applicable, where the $L$ point of the rock salt cell relates to the $\Gamma$ point in the conventional hexagonal cell. We construct an effective Hamiltonian of GeTe~(dRk) by fitting the first-principles data of the bulk unit cell presented in Sec.~\ref{sec:II2} near the $\Gamma$ point to our effective model. The parameter set is given in Tab.~\ref{tab:2}. Related parameters for a Hamiltonian that describes the GST225 (Kooi) phase are also given in Tab.~\ref{tab:2}.

\begin{figure}[t]
\begin{center}
\subfigure[]{\scalebox{0.8}{\includegraphics{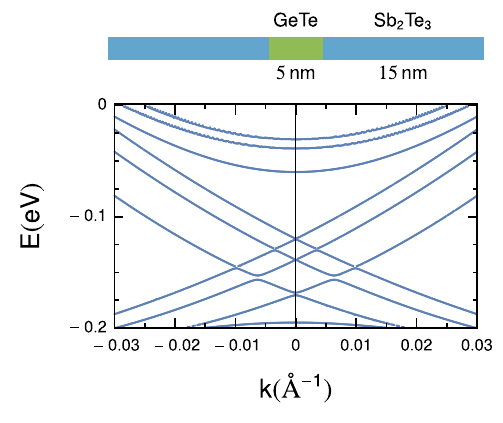}}\label{fig:5a}\qquad}
\subfigure[]{\scalebox{0.79}{\includegraphics{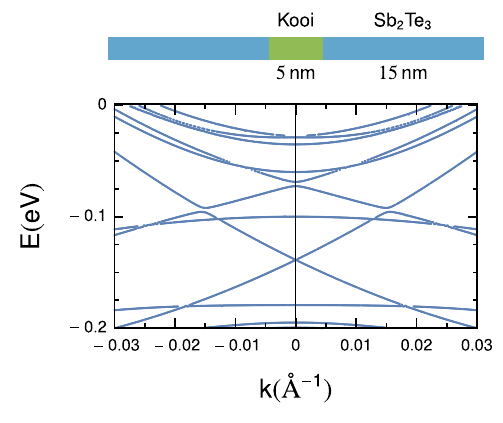}}\label{fig:5b}\qquad}
\caption{Calculated results for the band structure of an SGS tri-layer in the continuum model. (a) Band structure of an Sb$_2$Te$_3$/GeTe/Sb$_2$Te$_3$ tri-layer. (b) Band structure of an Sb$_2$Te$_3$/GST225/Sb$_2$Te$_3$ tri-layer. 
}
\label{fig:5}
\end{center}
\end{figure}
%++++++++++++++++++++++++++++++++++++++++

We now discuss the band structure as obtained from the envelope function formalism of an SGS tri-layer that consists of two outer Sb$_2$Te$_3$ layers and an embedded GeTe middle layer. 
% Results for a single Sb$_2$Te$_3$ slab are given in the Supplementary Material~\cite{SOM}.
Figure~\ref{fig:5a} shows results for the band structure near the $\Gamma$ point, where we choose an outer bulk-like Sb$_2$Te$_3$ layer of thickness $L_s=15$~nm, and an inner GeTe layer of thickness $L_g=5$~nm. While the surface states at the outer edges of the tri-layer are still localised at the slab energy of $-0.14$~eV, the inner Dirac states are shifted in energy, but no band gap opens at the $\Gamma$ point. Our numerical result is essentially unchanged for different values of the thickness $L_g$ of the inner GeTe layer, and a gap opening at the $\Gamma$ point is only observed for very small values $L_g < 2$~nm. This result suggests that (i) the interlayer coupling is completely suppressed even for moderate GeTe layers and (ii) that the perturbation of the GeTe wave function is sufficiently weak that the band structure of SG interfacial state is unchanged. This  is a very reasonable result considering the large offset of the GeTe valence band maximum and conduction band minimum compared to the Sb$_2$Te$_3$ layer: The VBM and CBM of Sb$_2$Te$_3$ are located within the band gap of GeTe and there are no GeTe states in the energy range of surface states. The interlayer coupling of the topological mode is thus suppressed.

As an alternative model, we examine an SGS structure in which the GeTe layer is replaced by GST225 (Kooi) with a thickness of $L_k=5$~nm. The results are shown in Fig.~\ref{fig:5b}. Interestingly, and in contrast to the previous case, we find a band gap opening at the $\Gamma$ point. The band gap appears to close at points away from the $\Gamma$ point, but this closed gap should be considered as an accidental degeneracy of the wave functions originating from the band structure of Sb$_2$Te$_3$ and GST225(Kooi), rather than any topological modes. However, the behaviour at the $\Gamma$ point indicates clearly that the microscopic structure of the NI region strongly influences the overall band structure. Within the continuum model we understand this as being a consequence of the reduced band offset between Sb$_2$Te$_3$ and GST225 compared to Sb$_2$Te$_3$ and simply GeTe which facilitates interlayer coupling of the topological modes. Further insights into the nature of the ETSs are obtained in the following section.

\section{FIRST-PRINCIPLES CALCULATIONS OF INTERFACIAL STATES IN SGS TRI-LAYER}\label{sec:IV}

The continuum model assumes that the Hamiltonian derived from the (homogeneous) bulk structure is applicable to the heterostructure junction. Although we mimicked the effect of physical intermixing of chemical species by introducing the GST225 layer, effects due to local electronic states and/or vacancies, which we refer to as ``chemical interactions'', are not represented. While charge transfer or charge accumulation by impurities at the SG interface is expected to be sufficiently small (as the electron affinity of both Sb$_2$Te$_3$ and GeTe is strong), the robustness of any topological mode against local fields arising from chemical interactions effect is not clear. In order to address this, in the following, we extract interfacial states of various thin SGS tri-layers directly by using NEGF-DFT. 

In our computational model of the tri-layer, the left- and right-hand side of Sb$_2$Te$_3$ is represented explicitly by $1\times1\times3$ unit cells, where the outermost cells are connected to the bulk by self-energy terms as discussed in Sec.~\ref{sec:II2}. Hence, different from the previous section, we can only consider the embedded interfacial state on the SG interface sides to analyze the topological mode, i.e., the intralayer-coupling is automatically eliminated. We consider three separate models, A, B, and C, with three separate structures for the NI part. Model A represents a sharp SG interface, Model B depicts a disordered SG interface, and Model C represents the situation where there is an intermixed region (GST225) at the SG interface. Based on our APT measurements in Section~\ref{sec:experiment}, Model C is the closest approximation of the system we consider here.\\

Model A: [(Sb$_2$Te$_3$)$_9$] /(GeTe)$_{3n}$/[(Sb$_2$Te$_3$)$_9$]  (ideal interface)
Model B: [(Sb$_2$Te$_3$)$_9$]/(Ge$_2$Te$_2$)(GeTe)$_{3(n-2)}$(Ge$_2$Te$_2$) /[(Sb$_2$Te$_3$)$_9$]  (disordered interface)

Model C: [(Sb$_2$Te$_3$)$_9$] /(GST225)$_m$(GeTe)$_{3(n-2)}$(GST225)$_m$[(Sb$_2$Te$_3$)$_9$] (realistic interface)\\

In our notation, [(Sb$_2$Te$_3$)$_3$]$_2$, for example, denotes a staking of the two (conventional hexagonal) unit cells of the Sb$_2$Te$_3$ crystal, i.e., it is a stacking of six Sb$_2$Te$_3$ quintuple monolayers (QLs). The intermediate layer in Model A is a GeTe~(dRk) layer. Here, the stacked numbers of (GeTe)$_3$ units, $n$, is taken as $n = 6$ (recall that a stacking of three GeTe monolayers is also the unit cell of the GeTe~(dRK) bulk crystal in the conventional hexagonal cell). A change in the interface structure is taken into account in modeM B, which contains a vacancy layer at the boundary of a GeTe~(dRk) phase. Here, the label (Ge$_2$Te$_2$) represents a vacancy layer that consists of a single (GeTe)$_3$ block. Finally, in Model C we introduce two unit cells of GST225 on either side of GeTe as an intermixing region, i.e., $m=2$. The outermost regions are connected to bulk Sb$_2$Te$_3$~(Rh). The 2D band dispersion of the interface is extracted by projecting the density of states (DOS) on the interface, and is exactly calculated from the Green’s function as a function of energy $E$ and wave vector $k_{||}$. We take $k$ to point along M-$\Gamma$-K line and the DOS was projected on each QL in the junction. We present the projected band structure at three separate positions: (i) the Sb$_2$Te$_3$-QL closest to the SG interface plane, (ii) the secondary neighboring QL, and (iii) the third QL in order to analyze the localization of topological mode. We labeled the above the three QLs as QL$^{(i)}$, QL$^{(ii)}$, and QL$^{(iii)}$, respectively.

\begin{figure}[t]
\begin{center}
\subfigure[]{
\scalebox{0.85}{\includegraphics{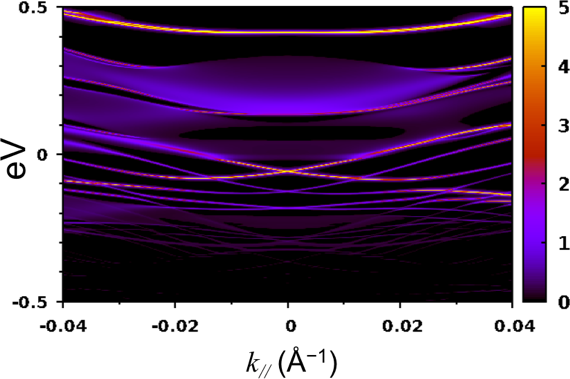}} \quad
\scalebox{0.85}{\includegraphics{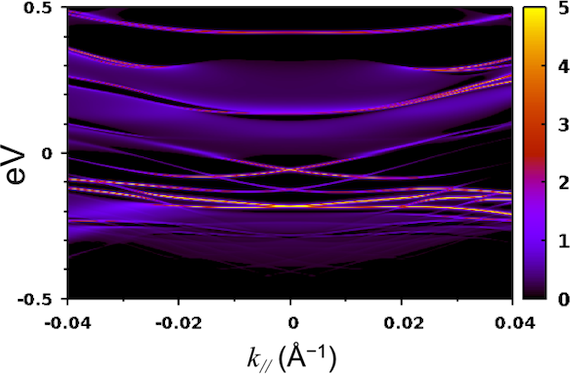}} \quad
\scalebox{0.85}{\includegraphics{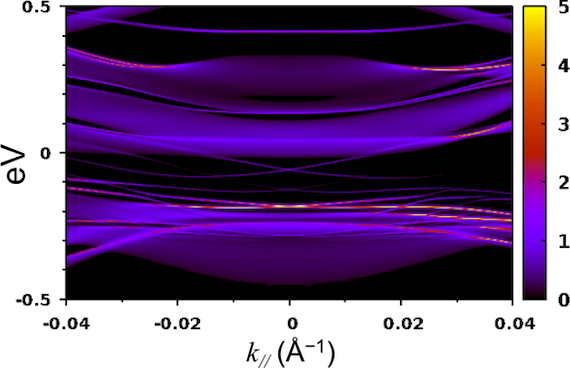}}
} \\
\subfigure[]{
\scalebox{0.85}{\includegraphics{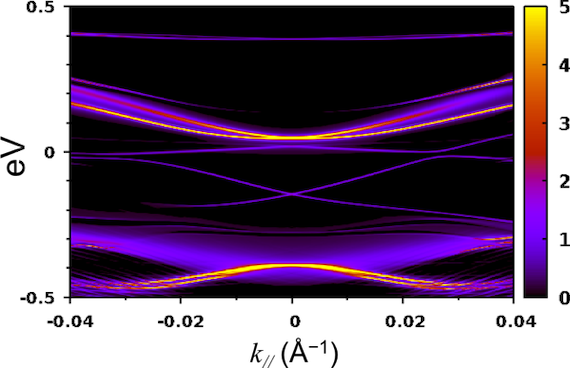}} \quad
\scalebox{0.85}{\includegraphics{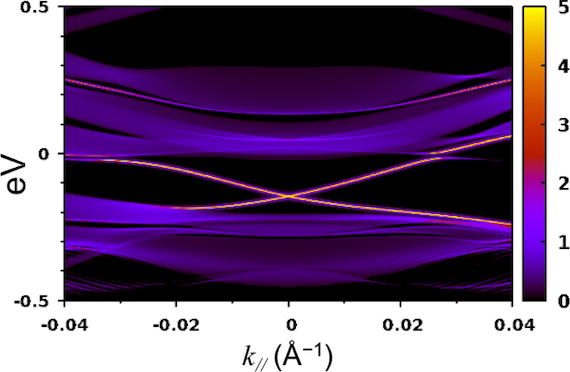}} \quad
\scalebox{0.85}{\includegraphics{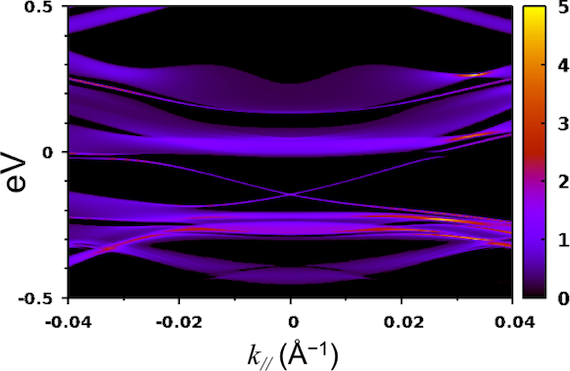}}
} \\
\subfigure[]{
\scalebox{0.85}{\includegraphics{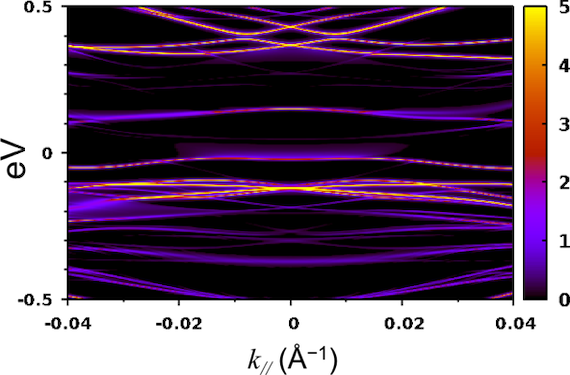}} \quad
\scalebox{0.85}{\includegraphics{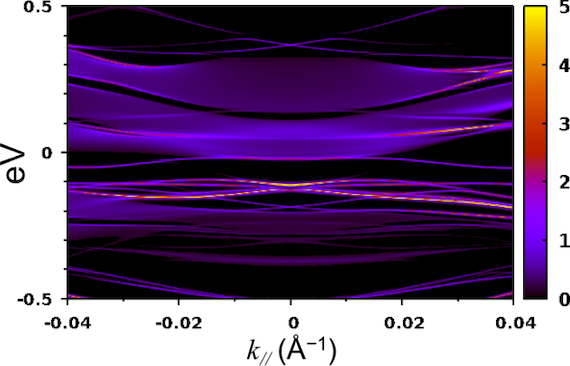}} \quad
\scalebox{0.85}{\includegraphics{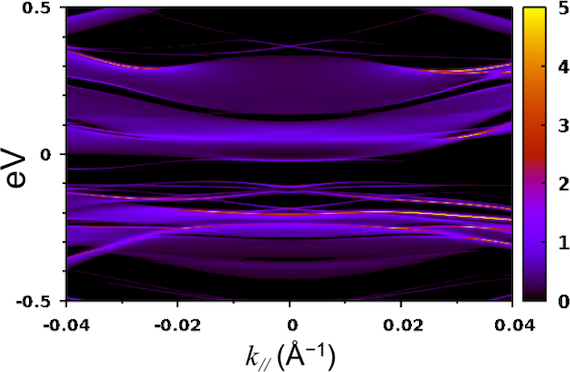}}
}
\caption{The extracted 2D band structures projected on a Sb$_2$Te$_3$ quintuple layer in the SGS tri-layer system by NEGF-DFT calculation. The Fermi level of bulk Sb$_2$Te$_3$ is set to zero. The zero of $k_{||}$ is at the $\Gamma$ point, and $k_{||}$ is positive along the $\Gamma$-K line and negative along the $\Gamma$-M line. Three SGS tri-layer structures denoted as model A, B, and C were examined (see Sec.~\ref{sec:IV} of the text for the definition), corresponding to panels (a), (b), and (c), respectively. The left column is the 2D band structure projected on QL$^{(i)}$, i.e., the Sb$_2$Te$_3$ quintuple layer nearest to the GeTe (or GST225) block. The middle and right columns show band structures projected on the second (QL$^{(ii)}$) and third (QL$^{(iii)}$) nearest quintuple layer, respectively. }
\label{fig:6}
\end{center}
\end{figure}

In Model A~(Fig.~\ref{fig:6}a) and Model B~(Fig.~\ref{fig:6}b), the extracted 2D band structure on QL$^{(iii)}$ is very similar to that of bulk Sb$_2$Te$_3$~(Rh) as given in Fig.~\ref{fig:5a}. Although a very weak spectral density coming from the DOS of QL$^{(ii)}$ is found, the electronic state in QL$^{(iii)}$ is essentially a bulk state for both Model A and Model B. In contrast, the projected band dispersion of QL$^{(i)}$ is more complicated and shows strong hybridization with the states of GeTe, i.e., the Sb$_2$Te$_3$ layer immediately adjacent to GeTe is strongly perturbed by chemical interactions. In the QL$^{(ii)}$ of Model B, we find a clear Dirac cone with a much stronger spectral density than in QL$^{(i)}$, and similar to the clean Sb$_2$Te$_3$ surface state. This is consistent with the results of Schubert~\textit{et al.}~\cite{Schubert2012} wherein it is found that strong disorder can shift the topological state away from the surface and into the material. Interestingly, for Model A we observe a Rashba-type split band rather than topological mode. While it is hard to gauge this unexpected result against experimental measurements as it represents an ideal interface, unlikely to occur in real systems, we note that this finding is comparable to that seen in Ref.~\cite{Menshov2015} where ETSs appear to have a Rashba-like character. 

These results lead to the following conclusions: first, the interfacial state characterised as the “surface” state of Sb$_2$Te$_3$ can be localised narrowly in the secondary neighboring Sb$_2$Te$_3$ monolayer in the junction; and second, the topological mode is not robust to chemical interaction even when the band offset is sufficiently large. By comparing models A and B, the existence of a vacancy layer at the SG interface induces a significant chemical interaction effect. We speculate that the local electric field due to a (GeTe)$_3$ block in the SG interface works as a built-in asymmetric external field and gives rise to a Rashba-type interfacial state even though the topological mode is protected by GeTe block containing a vacancy layer, i.e., Ge$_2$Te$_2$. 

Interestingly, in contrast to model A and model B, we found that 2D band on QL$^{(iii)}$ of model C (Fig.~\ref{fig:6}c) does not converge to that of bulk. The presence of GST225 opens large band gap and the band dispersion is more NI-like near the $\Gamma$ point. This result is also found in the continuum model (Fig.~\ref{fig:5b}), %and thus the interlayer coupling of the wave function across GST225 is more long-ranged than in the cases with GeTe or GeTe vacancy layer. 
although Model C represents a somewhat more realistic intermixing at the SG interface in which GST225 is only narrow sublayer of NI rather than the entire NI slab. % (GST225)$_m$[(GeTe)3]$_{(n-2)}$(GST225)$_m$, where $(m,n) = (2,6)$.
The absence of a bandgap in %According to the results of 
Model A and B suggests that the GeTe block is sufficiently thick to suppress interlayer coupling of topological modes from the two Sb$_2$Te$_3$ slabs. However, this changes dramatically in Model C wherein, for the same thickness of the NI slab, one finds a band gap. In other words, the large band gap at $\Gamma$ is a direct consequence of the existence of GST225. %result of local interaction (and hybridization) of each topological mode and electronic state narrowly localised on GST225. 

The origin of the band gap in Model C can be due to disorder~\cite{Schubert2012} or it may be due to \textit{local} interactions and/or hybridization of each topological mode with the electronic state narrowly localised on GST225. A third possibility is that it is due to \textit{long-range} coupling of ETSs on either side of the NI region as suggested by Nguyen~\textit{et al.}~\cite{nguyen16} We can rule out disorder as the cause since the first principles calculations are performed for ideal crystalline systems of which only Model B incorporates disorder. Based on the first principles results we cannot categorically rule out or posit either of the other options, i.e., local or long-range interactions. However, we can use the continuum model to conclude the following: it follows from Fig.~\ref{fig:5b} that if the NI region is fully GST225 then the band gap must be a consequence of long-range coupling of ETSs as disorder is not included in the model. The intermixed regions in our samples are $\approx 6 - 8$~nm thick (Fig.~3),%\ref{fig:APT})
 which is comparable to the NI thickness considered in Models A - C. Moreover, Fig.~3 also indicates a finite intermixing across the entire NI slab which indicates that the experimental system corresponds closely to the continuum model considered in Fig.~\ref{fig:5b}. This would then lead to the conclusion that if the intermixed GST225 region is sufficiently thick in the NI part, the topological modes can be made to couple by tuning ``bulk-like'' thickness of the NI in the SGS tri-layer. Thus, these results strongly suggest the role of long-ranged coupling of ETSs in the results %is offers a very plausible explanation for the results 
 observed by Nguyen~\textit{et al.}~\cite{nguyen16}.% since, experimentally, we find clear evidence of an intermixing region even in bulk-like SGS heterostructures (Figs.~1 and~\ref{APT}).

%++++++++++++++++++++++++++++++++++++++++
%++++++++++++++++++++++++++++++++++++++++

\section{CONCLUSIONS}

In this paper, we have studied the band structure of bulk-like heterostructures of Sb$_2$Te$_3$ and GeTe as a prototypical TI-NI system. We have focused on the interfacial region between these materials and shed light on the conditions under which the topological mode may or may not be present. We have laid particular emphasis on understanding realistic structures in which the interface is not perfect, but rather contains an intermixed phase of the parent compounds Sb$_2$Te$_3$ and GeTe. Our principal finding is that the presence of this intermediate phase serves to enhance the length over which topological modes may interact with each other. Importantly, we confirm experimentally that SGS heterostructures show a degree of intermixing, thereby underlining the relevance of our findings to the existing experimental literature on SGS systems. Of particular importance in this context is the experimental work by Nguyen~\textit{et al.}~\cite{nguyen16} which observes unexpectedly long-ranged interactions between topological modes in an SGS heterostructure. Our findings provide a natural explanation for these results and lay the foundation for future work wherein superlattices of bulk-like Sb$_2$Te$_3$ and GeTe layers can be used to deterministically produce different topological phases~\cite{burkov11}.

\section*{Acknowledgements}

This work was supported by EPSRC, UK, CREST, JST (Grant No. JPMJCR14F1 and JPMJCR18I4) and Peterhouse, Cambridge.

\section*{References}

\bibliography{bib}

\begin{thebibliography}{10}
\urlstyle{rm}
\expandafter\ifx\csname url\endcsname\relax
  \def\url#1{\texttt{#1}}\fi
\expandafter\ifx\csname urlprefix\endcsname\relax\def\urlprefix{URL }\fi
\expandafter\ifx\csname doiprefix\endcsname\relax\def\doiprefix{DOI: }\fi
\providecommand{\bibinfo}[2]{#2}
\providecommand{\eprint}[2][]{\url{#2}}

\bibitem{burkov11}
\bibinfo{author}{Burkov, A.~A.} \& \bibinfo{author}{Balents, L.}
\newblock \bibinfo{journal}{\bibinfo{title}{Weyl semimetal in a topological
  insulator multilayer}}.
\newblock {\emph{\JournalTitle{Phys. Rev. Lett.}}}
  \textbf{\bibinfo{volume}{107}}, \bibinfo{pages}{127205},
  \doiprefix\url{10.1103/PhysRevLett.107.127205} (\bibinfo{year}{2011}).

\bibitem{tsu68}
\bibinfo{author}{Tsu, R.}, \bibinfo{author}{Howard, W.~E.} \&
  \bibinfo{author}{Esaki, L.}
\newblock \bibinfo{journal}{\bibinfo{title}{Optical and electrical properties
  and band structure of $\mathrm{GeTe}$ and $\mathrm{SnTe}$}}.
\newblock {\emph{\JournalTitle{Phys. Rev.}}} \textbf{\bibinfo{volume}{172}},
  \bibinfo{pages}{779}, \doiprefix\url{10.1103/PhysRev.172.779}
  (\bibinfo{year}{1968}).

\bibitem{hein64}
\bibinfo{author}{Hein, R.~A.}, \bibinfo{author}{Gibson, J.~W.},
  \bibinfo{author}{Mazelsky, R.}, \bibinfo{author}{Miller, R.~C.} \&
  \bibinfo{author}{Hulm, J.~K.}
\newblock \bibinfo{journal}{\bibinfo{title}{Superconductivity in germanium
  telluride}}.
\newblock {\emph{\JournalTitle{Phys. Rev. Lett.}}}
  \textbf{\bibinfo{volume}{12}}, \bibinfo{pages}{320},
  \doiprefix\url{10.1103/PhysRevLett.12.320} (\bibinfo{year}{1964}).

\bibitem{narayan16}
\bibinfo{author}{Narayan, V.}, \bibinfo{author}{Nguyen, T.-A.},
  \bibinfo{author}{Mansell, R.}, \bibinfo{author}{Ritchie, D.} \&
  \bibinfo{author}{Mussler, G.}
\newblock \bibinfo{journal}{\bibinfo{title}{Interplay of spin--orbit coupling
  and superconducting correlations in germanium telluride thin films}}.
\newblock {\emph{\JournalTitle{physica status solidi (RRL) -- Rapid Research
  Letters}}} \textbf{\bibinfo{volume}{10}}, \bibinfo{pages}{253},
  \doiprefix\url{10.1002/pssr.201510430} (\bibinfo{year}{2016}).

\bibitem{narayan19}
\bibinfo{author}{Narayan, V.} \emph{et~al.}
\newblock \bibinfo{journal}{\bibinfo{title}{Long-lived nonequilibrium
  superconductivity in a noncentrosymmetric rashba semiconductor}}.
\newblock {\emph{\JournalTitle{Phys. Rev. B}}} \textbf{\bibinfo{volume}{100}},
  \bibinfo{pages}{024504}, \doiprefix\url{10.1103/PhysRevB.100.024504}
  (\bibinfo{year}{2019}).

\bibitem{hiseh09}
\bibinfo{author}{Hsieh, D.} \emph{et~al.}
\newblock \bibinfo{journal}{\bibinfo{title}{Observation of
  time-reversal-protected single-dirac-cone topological-insulator states in
  $\mathrm{Bi}_{2}\mathrm{Te}_{3}$ and $\mathrm{Sb}_{2}\mathrm{Te}_{3}$}}.
\newblock {\emph{\JournalTitle{Phys. Rev. Lett.}}}
  \textbf{\bibinfo{volume}{103}}, \bibinfo{pages}{146401},
  \doiprefix\url{10.1103/PhysRevLett.103.146401} (\bibinfo{year}{2009}).

\bibitem{zhang09}
\bibinfo{author}{Zhang, H.} \emph{et~al.}
\newblock \bibinfo{journal}{\bibinfo{title}{Topological insulators in
  $\mathrm{Bi}_2\mathrm{Se}_{3}$, $\mathrm{Bi}_{2}\mathrm{Te}_{3}$ and
  $\mathrm{Sb}_{2}\mathrm{Te}_{3}$ with a single dirac cone on the surface}}.
\newblock {\emph{\JournalTitle{Nature Physics}}} \textbf{\bibinfo{volume}{5}},
  \bibinfo{pages}{438}, \doiprefix\url{10.1038/nphys1270}
  (\bibinfo{year}{2009}).

\bibitem{raoux14}
\bibinfo{author}{Raoux, S.}, \bibinfo{author}{Xiong, F.},
  \bibinfo{author}{Wuttig, M.} \& \bibinfo{author}{Pop, E.}
\newblock \bibinfo{journal}{\bibinfo{title}{{Phase change materials and phase
  change memory}}}.
\newblock {\emph{\JournalTitle{MRS Bulletin}}} \textbf{\bibinfo{volume}{39}},
  \bibinfo{pages}{703}, \doiprefix\url{10.1557/mrs.2014.139}
  (\bibinfo{year}{2014}).

\bibitem{kato06}
\bibinfo{author}{Kato, N.} \emph{et~al.}
\newblock \bibinfo{journal}{\bibinfo{title}{$\mathrm{Ge}\mathrm{S}_2$/metal
  thin film bilayered structures as write-once-type optical recording
  materials}}.
\newblock {\emph{\JournalTitle{Journal of Applied Physics}}}
  \textbf{\bibinfo{volume}{100}}, \bibinfo{pages}{113115},
  \doiprefix\url{10.1063/1.2398556} (\bibinfo{year}{2006}).

\bibitem{yamada87}
\bibinfo{author}{Yamada, N.} \emph{et~al.}
\newblock \bibinfo{journal}{\bibinfo{title}{{High Speed Overwritable Phase
  Change Optical Disk Material}}}.
\newblock {\emph{\JournalTitle{Japanese Journal of Applied Physics}}}
  \textbf{\bibinfo{volume}{26}}, \bibinfo{pages}{61},
  \doiprefix\url{10.7567/jjaps.26s4.61} (\bibinfo{year}{1987}).

\bibitem{simpson11}
\bibinfo{author}{Simpson, R.~E.} \emph{et~al.}
\newblock \bibinfo{journal}{\bibinfo{title}{Interfacial phase-change memory}}.
\newblock {\emph{\JournalTitle{Nature Nanotechnology}}}
  \textbf{\bibinfo{volume}{6}}, \bibinfo{pages}{501} (\bibinfo{year}{2011}).

\bibitem{chen86}
\bibinfo{author}{Chen, M.}, \bibinfo{author}{Rubin, K.~A.} \&
  \bibinfo{author}{Barton, R.~W.}
\newblock \bibinfo{journal}{\bibinfo{title}{Compound materials for reversible,
  phase‐change optical data storage}}.
\newblock {\emph{\JournalTitle{Applied Physics Letters}}}
  \textbf{\bibinfo{volume}{49}}, \bibinfo{pages}{502},
  \doiprefix\url{10.1063/1.97617} (\bibinfo{year}{1986}).

\bibitem{inoue19}
\bibinfo{author}{Inoue, N.} \& \bibinfo{author}{Nakamura, H.}
\newblock \bibinfo{journal}{\bibinfo{title}{Structural transition pathway and
  bipolar switching of the gete--sb2te3 superlattice as interfacial
  phase-change memory}}.
\newblock {\emph{\JournalTitle{Faraday Discuss.}}}
  \textbf{\bibinfo{volume}{213}}, \bibinfo{pages}{303},
  \doiprefix\url{10.1039/C8FD00093J} (\bibinfo{year}{2019}).

\bibitem{tominaga14}
\bibinfo{author}{Tominaga, J.}, \bibinfo{author}{Kolobov, A.~V.},
  \bibinfo{author}{Fons, P.}, \bibinfo{author}{Nakano, T.} \&
  \bibinfo{author}{Murakami, S.}
\newblock \bibinfo{journal}{\bibinfo{title}{Ferroelectric order control of the
  dirac-semimetal phase in gete-sb2te3 superlattices}}.
\newblock {\emph{\JournalTitle{Advanced Materials Interfaces}}}
  \textbf{\bibinfo{volume}{1}}, \bibinfo{pages}{1300027},
  \doiprefix\url{10.1002/admi.201300027} (\bibinfo{year}{2014}).

\bibitem{tominaga15}
\bibinfo{author}{Tominaga, J.} \emph{et~al.}
\newblock \bibinfo{journal}{\bibinfo{title}{Giant multiferroic effects in
  topological gete-$\mathrm{Sb}_2\mathrm{Te}_3$ superlattices}}.
\newblock {\emph{\JournalTitle{Science and Technology of Advanced Materials}}}
  \textbf{\bibinfo{volume}{16}}, \bibinfo{pages}{014402},
  \doiprefix\url{10.1088/1468-6996/16/1/014402} (\bibinfo{year}{2015}).

\bibitem{ibarra18}
\bibinfo{author}{Ibarra-Hern\'andez, W.} \& \bibinfo{author}{Raty, J.-Y.}
\newblock \bibinfo{journal}{\bibinfo{title}{Ab initio density functional theory
  study of the electronic, dynamic, and thermoelectric properties of the
  crystalline pseudobinary chalcogenide
  $\text{(GeTe)}_{x}/(\mathrm{Sb}_{2}\mathrm{Te}_{3})\phantom{\rule{0.28em}{0ex}}(x=1,\phantom{\rule{0.28em}{0ex}}2,\phantom{\rule{0.28em}{0ex}}3)$}}.
\newblock {\emph{\JournalTitle{Phys. Rev. B}}} \textbf{\bibinfo{volume}{97}},
  \bibinfo{pages}{245205}, \doiprefix\url{10.1103/PhysRevB.97.245205}
  (\bibinfo{year}{2018}).

\bibitem{halasz12}
\bibinfo{author}{Hal\'asz, G.~B.} \& \bibinfo{author}{Balents, L.}
\newblock \bibinfo{journal}{\bibinfo{title}{Time-reversal invariant realization
  of the $\mathrm{Weyl}$ semimetal phase}}.
\newblock {\emph{\JournalTitle{Phys. Rev. B}}} \textbf{\bibinfo{volume}{85}},
  \bibinfo{pages}{035103}, \doiprefix\url{10.1103/PhysRevB.85.035103}
  (\bibinfo{year}{2012}).

\bibitem{sa12}
\bibinfo{author}{Sa, B.}, \bibinfo{author}{Zhou, J.}, \bibinfo{author}{Sun,
  Z.}, \bibinfo{author}{Tominaga, J.} \& \bibinfo{author}{Ahuja, R.}
\newblock \bibinfo{journal}{\bibinfo{title}{Topological insulating in
  $\mathrm{GeTe}/\mathrm{Sb}_{2}\mathrm{Te}_{3}$ phase-change superlattice}}.
\newblock {\emph{\JournalTitle{Phys. Rev. Lett.}}}
  \textbf{\bibinfo{volume}{109}}, \bibinfo{pages}{096802},
  \doiprefix\url{10.1103/PhysRevLett.109.096802} (\bibinfo{year}{2012}).

\bibitem{nakamura17}
\bibinfo{author}{Nakamura, H.} \emph{et~al.}
\newblock \bibinfo{journal}{\bibinfo{title}{Resistive switching mechanism of
  ${\text{gete}}$ gete--sb2te3 interfacial phase change memory and topological
  properties of embedded two-dimensional states}}.
\newblock {\emph{\JournalTitle{Nanoscale}}} \textbf{\bibinfo{volume}{9}},
  \bibinfo{pages}{9386}, \doiprefix\url{10.1039/C7NR03495D}
  (\bibinfo{year}{2017}).

\bibitem{bang16}
\bibinfo{author}{Bang, D.}, \bibinfo{author}{Awano, H.},
  \bibinfo{author}{Saito, Y.} \& \bibinfo{author}{Tominaga, J.}
\newblock \bibinfo{journal}{\bibinfo{title}{Temperature dependence of
  magneto-optical kerr signal in $\mathrm{GeTe}−sb_2te_3$ topological
  superlattice}}.
\newblock {\emph{\JournalTitle{AIP Advances}}} \textbf{\bibinfo{volume}{6}},
  \bibinfo{pages}{055810}, \doiprefix\url{10.1063/1.4943152}
  (\bibinfo{year}{2016}).

\bibitem{qian16}
\bibinfo{author}{Qian, H.} \emph{et~al.}
\newblock \bibinfo{journal}{\bibinfo{title}{Low work function of crystalline
  {GeTe}/sb2te3superlattice-like films induced by te dangling bonds}}.
\newblock {\emph{\JournalTitle{Journal of Physics D: Applied Physics}}}
  \textbf{\bibinfo{volume}{49}}, \bibinfo{pages}{495302},
  \doiprefix\url{10.1088/0022-3727/49/49/495302} (\bibinfo{year}{2016}).

\bibitem{Momand2015}
\bibinfo{author}{Momand, J.} \emph{et~al.}
\newblock \bibinfo{journal}{\bibinfo{title}{Interface formation of two- and
  three-dimensionally bonded materials in the case of gete–sb2te3
  superlattices}}.
\newblock {\emph{\JournalTitle{Nanoscale}}} \textbf{\bibinfo{volume}{7}},
  \bibinfo{pages}{19136--19143}, \doiprefix\url{10.1039/C5NR04530D}
  (\bibinfo{year}{2015}).

\bibitem{kim07}
\bibinfo{author}{Kim, J.-J.} \emph{et~al.}
\newblock \bibinfo{journal}{\bibinfo{title}{Electronic structure of amorphous
  and crystalline
  ${(\mathrm{Ge}\mathrm{Te})}_{1\ensuremath{-}x}{({\mathrm{Sb}}_{2}{\mathrm{Te}}_{3})}_{x}$
  investigated using hard x-ray photoemission spectroscopy}}.
\newblock {\emph{\JournalTitle{Phys. Rev. B}}} \textbf{\bibinfo{volume}{76}},
  \bibinfo{pages}{115124}, \doiprefix\url{10.1103/PhysRevB.76.115124}
  (\bibinfo{year}{2007}).

\bibitem{kraut80}
\bibinfo{author}{Kraut, E.~A.}, \bibinfo{author}{Grant, R.~W.},
  \bibinfo{author}{Waldrop, J.~R.} \& \bibinfo{author}{Kowalczyk, S.~P.}
\newblock \bibinfo{journal}{\bibinfo{title}{Precise determination of the
  valence-band edge in x-ray photoemission spectra: Application to measurement
  of semiconductor interface potentials}}.
\newblock {\emph{\JournalTitle{Phys. Rev. Lett.}}}
  \textbf{\bibinfo{volume}{44}}, \bibinfo{pages}{1620},
  \doiprefix\url{10.1103/PhysRevLett.44.1620} (\bibinfo{year}{1980}).

\bibitem{fang11}
\bibinfo{author}{Fang, L. W.-W.} \emph{et~al.}
\newblock \bibinfo{journal}{\bibinfo{title}{Band offsets between sio2 and phase
  change materials in the (gete)x(sb2te3)1−x pseudobinary system}}.
\newblock {\emph{\JournalTitle{Applied Physics Letters}}}
  \textbf{\bibinfo{volume}{98}}, \bibinfo{pages}{132103},
  \doiprefix\url{10.1063/1.3573787} (\bibinfo{year}{2011}).

\bibitem{nguyen16}
\bibinfo{author}{Nguyen, T.-A.} \emph{et~al.}
\newblock \bibinfo{journal}{\bibinfo{title}{Topological states and phase
  transitions in sb2te3-gete multilayers}}.
\newblock {\emph{\JournalTitle{Scientific Reports}}}
  \textbf{\bibinfo{volume}{6}}, \bibinfo{pages}{27716},
  \doiprefix\url{10.1038/srep27716} (\bibinfo{year}{2016}).

\bibitem{blavett93}
\bibinfo{author}{Blavette, D.}, \bibinfo{author}{Bostel, A.},
  \bibinfo{author}{Sarrau, J.~M.}, \bibinfo{author}{Deconihout, B.} \&
  \bibinfo{author}{Menand, A.}
\newblock \bibinfo{journal}{\bibinfo{title}{An atom probe for three-dimensional
  tomography}}.
\newblock {\emph{\JournalTitle{Nature}}} \textbf{\bibinfo{volume}{363}},
  \bibinfo{pages}{432--435}, \doiprefix\url{https://doi.org/10.1038/363432a0}
  (\bibinfo{year}{1993}).

\bibitem{bas95}
\bibinfo{author}{Bas, P.}, \bibinfo{author}{Bostel, A.},
  \bibinfo{author}{Deconihout, B.} \& \bibinfo{author}{Blavette, D.}
\newblock \bibinfo{journal}{\bibinfo{title}{A general protocol for the
  reconstruction of 3d atom probe data}}.
\newblock {\emph{\JournalTitle{Appl. Surf. Sci.}}} \textbf{\bibinfo{volume}{87
  -- 88}}, \bibinfo{pages}{298--304},
  \doiprefix\url{https://doi.org/10.1016/0169-4332(94)00561-3}
  (\bibinfo{year}{1995}).

\bibitem{SOM}
\bibinfo{journal}{\bibinfo{title}{Supplemental material}}.
\newblock {\emph{\JournalTitle{Sci. Rep.}}} .

\bibitem{soler02}
\bibinfo{author}{Soler, J.~M.} \emph{et~al.}
\newblock \bibinfo{journal}{\bibinfo{title}{The {SIESTA} method forab
  initioorder-nmaterials simulation}}.
\newblock {\emph{\JournalTitle{Journal of Physics: Condensed Matter}}}
  \textbf{\bibinfo{volume}{14}}, \bibinfo{pages}{2745},
  \doiprefix\url{10.1088/0953-8984/14/11/302} (\bibinfo{year}{2002}).

\bibitem{rocha06}
\bibinfo{author}{Rocha, A.~R.} \emph{et~al.}
\newblock \bibinfo{journal}{\bibinfo{title}{Spin and molecular electronics in
  atomically generated orbital landscapes}}.
\newblock {\emph{\JournalTitle{Phys. Rev. B}}} \textbf{\bibinfo{volume}{73}},
  \bibinfo{pages}{085414}, \doiprefix\url{10.1103/PhysRevB.73.085414}
  (\bibinfo{year}{2006}).

\bibitem{sun06}
\bibinfo{author}{Sun, Z.}, \bibinfo{author}{Zhou, J.} \&
  \bibinfo{author}{Ahuja, R.}
\newblock \bibinfo{journal}{\bibinfo{title}{Structure of phase change materials
  for data storage}}.
\newblock {\emph{\JournalTitle{Phys. Rev. Lett.}}}
  \textbf{\bibinfo{volume}{96}}, \bibinfo{pages}{055507},
  \doiprefix\url{10.1103/PhysRevLett.96.055507} (\bibinfo{year}{2006}).

\bibitem{ohyanagi14}
\bibinfo{author}{Ohyanagi, T.} \emph{et~al.}
\newblock \bibinfo{journal}{\bibinfo{title}{Gete sequences in superlattice
  phase change memories and their electrical characteristics}}.
\newblock {\emph{\JournalTitle{Applied Physics Letters}}}
  \textbf{\bibinfo{volume}{104}}, \bibinfo{pages}{252106},
  \doiprefix\url{10.1063/1.4886119} (\bibinfo{year}{2014}).

\bibitem{disante13}
\bibinfo{author}{Di~Sante, D.}, \bibinfo{author}{Barone, P.},
  \bibinfo{author}{Bertacco, R.} \& \bibinfo{author}{Picozzi, S.}
\newblock \bibinfo{journal}{\bibinfo{title}{Electric control of the giant
  rashba effect in bulk gete}}.
\newblock {\emph{\JournalTitle{Advanced Materials}}}
  \textbf{\bibinfo{volume}{25}}, \bibinfo{pages}{509},
  \doiprefix\url{10.1002/adma.201203199} (\bibinfo{year}{2013}).

\bibitem{dasilva08}
\bibinfo{author}{Da~Silva, J. L.~F.}, \bibinfo{author}{Walsh, A.} \&
  \bibinfo{author}{Lee, H.}
\newblock \bibinfo{journal}{\bibinfo{title}{Insights into the structure of the
  stable and metastable
  ${(\text{GeTe})}_{m}{({\text{Sb}}_{2}{\text{Te}}_{3})}_{n}$ compounds}}.
\newblock {\emph{\JournalTitle{Phys. Rev. B}}} \textbf{\bibinfo{volume}{78}},
  \bibinfo{pages}{224111}, \doiprefix\url{10.1103/PhysRevB.78.224111}
  (\bibinfo{year}{2008}).

\bibitem{chang1966}
\bibinfo{author}{{Chang}, L.~L.}, \bibinfo{author}{{Stiles}, P.~J.} \&
  \bibinfo{author}{{Esaki}, L.}
\newblock \bibinfo{journal}{\bibinfo{title}{Electron barriers in al-al2o3-snte
  and al-al2o3-gete tunnel junctions}}.
\newblock {\emph{\JournalTitle{IBM Journal of Research and Development}}}
  \textbf{\bibinfo{volume}{10}}, \bibinfo{pages}{484--486}
  (\bibinfo{year}{1966}).

\bibitem{Esaki1966}
\bibinfo{author}{Esaki, L.} \& \bibinfo{author}{Stiles, P.~J.}
\newblock \bibinfo{journal}{\bibinfo{title}{New type of negative resistance in
  barrier tunneling}}.
\newblock {\emph{\JournalTitle{Phys. Rev. Lett.}}}
  \textbf{\bibinfo{volume}{16}}, \bibinfo{pages}{1108--1111},
  \doiprefix\url{10.1103/PhysRevLett.16.1108} (\bibinfo{year}{1966}).

\bibitem{Palaz2017}
\bibinfo{author}{Palaz, S.}, \bibinfo{author}{Koc, H.},
  \bibinfo{author}{Mamedov, A.~M.} \& \bibinfo{author}{Ozbay, E.}
\newblock \bibinfo{journal}{\bibinfo{title}{Topological insulators: Electronic
  band structure and spectroscopy}}.
\newblock {\emph{\JournalTitle{{IOP} Conference Series: Materials Science and
  Engineering}}} \textbf{\bibinfo{volume}{175}}, \bibinfo{pages}{012004},
  \doiprefix\url{10.1088/1757-899x/175/1/012004} (\bibinfo{year}{2017}).

\bibitem{kooi02}
\bibinfo{author}{Kooi, B.~J.} \& \bibinfo{author}{De~Hosson, J. T.~M.}
\newblock \bibinfo{journal}{\bibinfo{title}{Electron diffraction and
  high-resolution transmission electron microscopy of the high temperature
  crystal structures of gexsb2te3+x (x=1,2,3) phase change material}}.
\newblock {\emph{\JournalTitle{Journal of Applied Physics}}}
  \textbf{\bibinfo{volume}{92}}, \bibinfo{pages}{3584},
  \doiprefix\url{10.1063/1.1502915} (\bibinfo{year}{2002}).

\bibitem{bastard86}
\bibinfo{author}{{Bastard}, G.} \& \bibinfo{author}{{Brum}, J.}
\newblock \bibinfo{journal}{\bibinfo{title}{Electronic states in semiconductor
  heterostructures}}.
\newblock {\emph{\JournalTitle{IEEE Journal of Quantum Electronics}}}
  \textbf{\bibinfo{volume}{22}}, \bibinfo{pages}{1625},
  \doiprefix\url{10.1109/JQE.1986.1073186} (\bibinfo{year}{1986}).

\bibitem{liu10}
\bibinfo{author}{Liu, C.-X.} \emph{et~al.}
\newblock \bibinfo{journal}{\bibinfo{title}{Model hamiltonian for topological
  insulators}}.
\newblock {\emph{\JournalTitle{Phys. Rev. B}}} \textbf{\bibinfo{volume}{82}},
  \bibinfo{pages}{045122}, \doiprefix\url{10.1103/PhysRevB.82.045122}
  (\bibinfo{year}{2010}).

\bibitem{liebmann16}
\bibinfo{author}{Liebmann, M.} \emph{et~al.}
\newblock \bibinfo{journal}{\bibinfo{title}{Giant rashba-type spin splitting in
  ferroelectric gete(111)}}.
\newblock {\emph{\JournalTitle{Advanced Materials}}}
  \textbf{\bibinfo{volume}{28}}, \bibinfo{pages}{560},
  \doiprefix\url{10.1002/adma.201503459} (\bibinfo{year}{2016}).

\bibitem{Schubert2012}
\bibinfo{author}{Schubert, G.}, \bibinfo{author}{Fehske, H.},
  \bibinfo{author}{Fritz, L.} \& \bibinfo{author}{Vojta, M.}
\newblock \bibinfo{journal}{\bibinfo{title}{Fate of topological-insulator
  surface states under strong disorder}}.
\newblock {\emph{\JournalTitle{Phys. Rev. B}}} \textbf{\bibinfo{volume}{85}},
  \bibinfo{pages}{201105}, \doiprefix\url{10.1103/PhysRevB.85.201105}
  (\bibinfo{year}{2012}).

\bibitem{Menshov2015}
\bibinfo{author}{Men'shov, V.~N.}, \bibinfo{author}{Tugushev, V.~V.},
  \bibinfo{author}{Eremeev, S.~V.}, \bibinfo{author}{Echenique, P.~M.} \&
  \bibinfo{author}{Chulkov, E.~V.}
\newblock \bibinfo{journal}{\bibinfo{title}{Band bending driven evolution of
  the bound electron states at the interface between a three-dimensional
  topological insulator and a three-dimensional normal insulator}}.
\newblock {\emph{\JournalTitle{Phys. Rev. B}}} \textbf{\bibinfo{volume}{91}},
  \bibinfo{pages}{075307}, \doiprefix\url{10.1103/PhysRevB.91.075307}
  (\bibinfo{year}{2015}).

\end{thebibliography}

\section*{Author contribution statement}
This study was designed by H.N and V.N. H.N. and N.I. carried out the first-principles calculations and analyzed the data, J.H. constructed the continuum model and associated calculations, S.K. and P.M.K. did the APT measurements, G.M. and D.G. grew the samples, and V.N. did the XPS measurements. H.N., J.H., and V.N. wrote the paper with inputs from all the others.

\end{document}